\documentclass{article}

    \PassOptionsToPackage{numbers, compress}{natbib}

    \usepackage[preprint]{neurips_2024}

\usepackage[utf8]{inputenc} 
\usepackage[T1]{fontenc}    
\usepackage{hyperref}       
\usepackage{url}            
\usepackage{booktabs}       
\usepackage{amsfonts}       
\usepackage{nicefrac}       
\usepackage{microtype}      
\usepackage{xcolor}         
\usepackage{amsmath}
\usepackage{amssymb}
\usepackage{dsfont}
\usepackage{mathtools}
\usepackage{amsthm}
\usepackage{tabularx}
\usepackage{enumitem}
\usepackage{booktabs}
\usepackage{siunitx}

\usepackage{color}
\usepackage{float}
\usepackage{graphicx}%
\usepackage{amsmath}
\usepackage{adjustbox}
\usepackage{subfigure}
\usepackage{wrapfig}  
\usepackage{algorithm}%
\usepackage{algorithmicx}%
\usepackage{algpseudocode}%
\usepackage{xcolor}

\title{Advancing Symbolic Discovery on Unsupervised Data: 
A Pre-training Framework for Non-degenerate Implicit Equation Discovery}

\author{
Yufei~Kuang\textsuperscript{1},\,
Jie~Wang\textsuperscript{1}\thanks{Corresponding author. Email: jiewangx@ustc.edu.cn.}\,\,,\,
Haotong~Huang\textsuperscript{1},\,
Mingxuan~Ye\textsuperscript{1},\,
Fangzhou~Zhu\textsuperscript{2},\,
Xijun~Li\textsuperscript{3},\,\\
\textbf{Jianye}~\textbf{Hao}\textsuperscript{2,4},\,
\textbf{Feng}~\textbf{Wu}\textsuperscript{1}\\
\textsuperscript{1}MoE Key Laboratory of
Brain-inspired Intelligent Perception and Cognition,\\
University of Science and Technology of China\\
\textsuperscript{2} Noah’s Ark Lab, Huawei Technologies \\
\textsuperscript{3} Shanghai Jiao Tong University \\
\textsuperscript{4} Tianjin University
}

\begin{document}

\maketitle

\begin{abstract}
Symbolic regression (SR)---which learns symbolic equations to describe the underlying relation from input-output pairs---is widely used for scientific discovery. 
However, a rich set of scientific data from the real world (e.g., particle trajectories and astrophysics) are typically unsupervised, devoid of explicit input-output pairs. 
In this paper, we focus on symbolic implicit equation discovery, which aims to discover the mathematical relation from \textit{unsupervised} data that follows an implicit equation $f(\mathbf{x}) =0$. 
However, due to the dense distribution of degenerate solutions (e.g., $f(\mathbf{x})=x_i-x_i$) in the discrete search space, 
most existing SR approaches customized for this task fail to achieve satisfactory performance. 
To tackle this problem, we introduce a novel pre-training framework---namely, \textbf{P}re-trained neural symbolic model for \textbf{I}mplicit \textbf{E}quation (PIE)---to discover implicit equations from unsupervised data. 
The core idea is that, we formulate the implicit equation discovery on unsupervised scientific data as a translation task and utilize the prior learned from the pre-training dataset to infer  non-degenerate skeletons of the underlying relation end-to-end. 
Extensive experiments shows that, leveraging the prior from a pre-trained language model, PIE \textit{effectively} tackles the problem of degenerate solutions and \textit{significantly} outperforms \textit{all} the existing SR approaches. 
PIE shows an encouraging step towards general scientific discovery on unsupervised data.

\end{abstract}

\section{Introduction}

Given a set of \textit{unsupervised} data points sampled from the elliptical equation $\frac{x_1^2}{a}+\frac{x_2^2}{b}-1=0$, how can we recover the underlying equation from the given data? 
This fundamental problem represents a typical class of scientific discovery, which is widely studied in research fields like particle trajectories, astrophysics, and atmospheric dynamics \citep{ai4science,db-fem,cl-fem}. Formally, in these tasks, researchers aim to discover the mathematical relation from unsupervised data that follows an implicit equation $f(\mathbf{x}) =0$ \citep{math}. 

Machine learning (ML) is widely regarded as a powerful tool for automatically learning patterns on unsupervised data \citep{ai4science,survey-unsupervised}. 
However, many ML approaches employ black-box neural networks (NNs) as function approximators for prediction-only purpose, lacking transparency in explaining the learned relation \citep{symb,nesymbres,symb-graph-1}. 
In contrast, symbolic regression (SR)---which learns \textit{interpretable} symbolic equations to describe the relation from input-output pairs---is a powerful tool for scientific discovery \citep{symb-graph-1,symb-graph-2}. 
Thus, SR is a natural approach for discovering implicit equations from unsupervised data.

\begin{figure}[t]
    \centering
    \subfigure[Particle trajectories.]{
        \label{fig: example1}
        \includegraphics[width=0.33\textwidth]{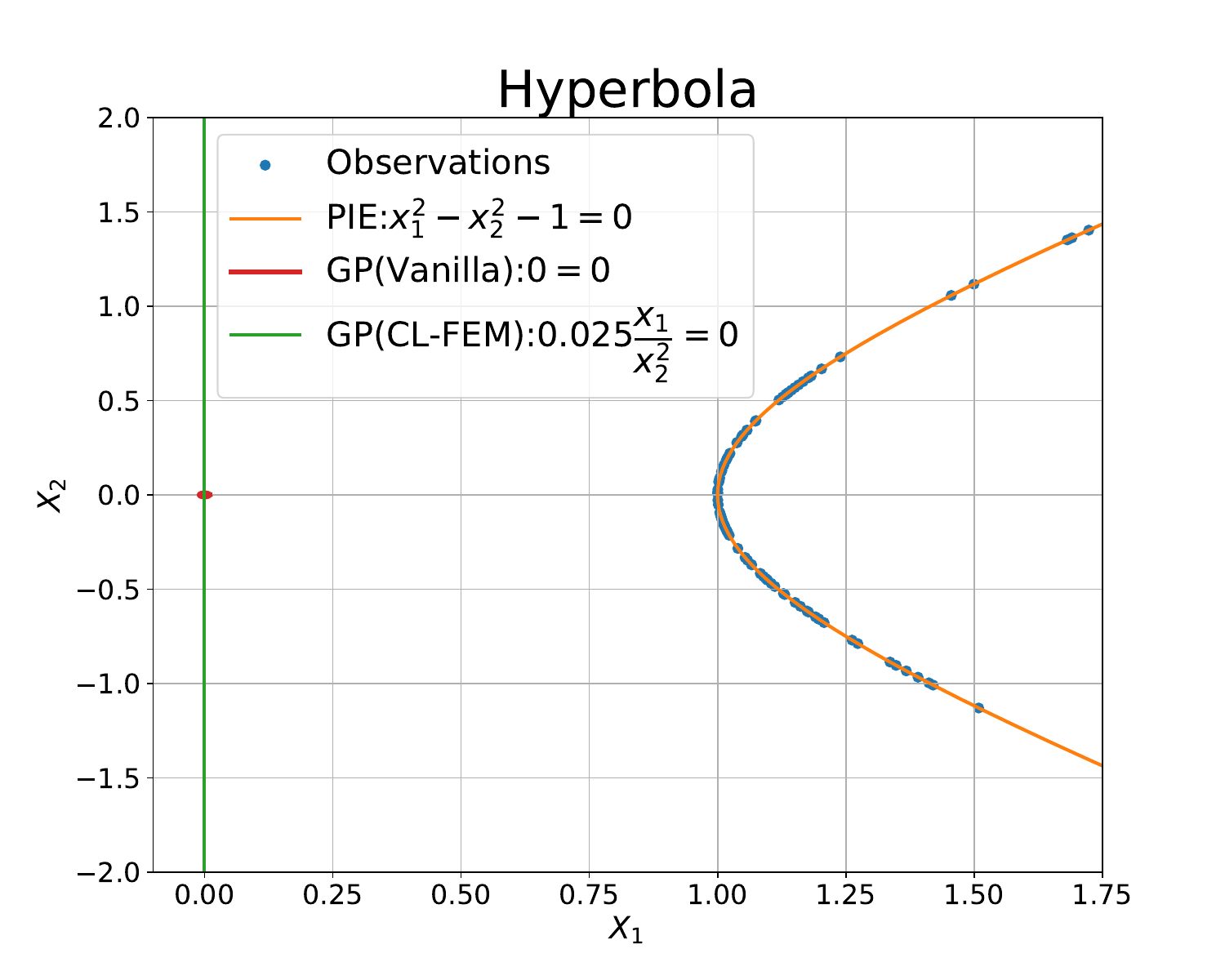}}
    \subfigure[Celestial motion.]{
        \label{fig: example2}
        \includegraphics[width=0.33\textwidth]{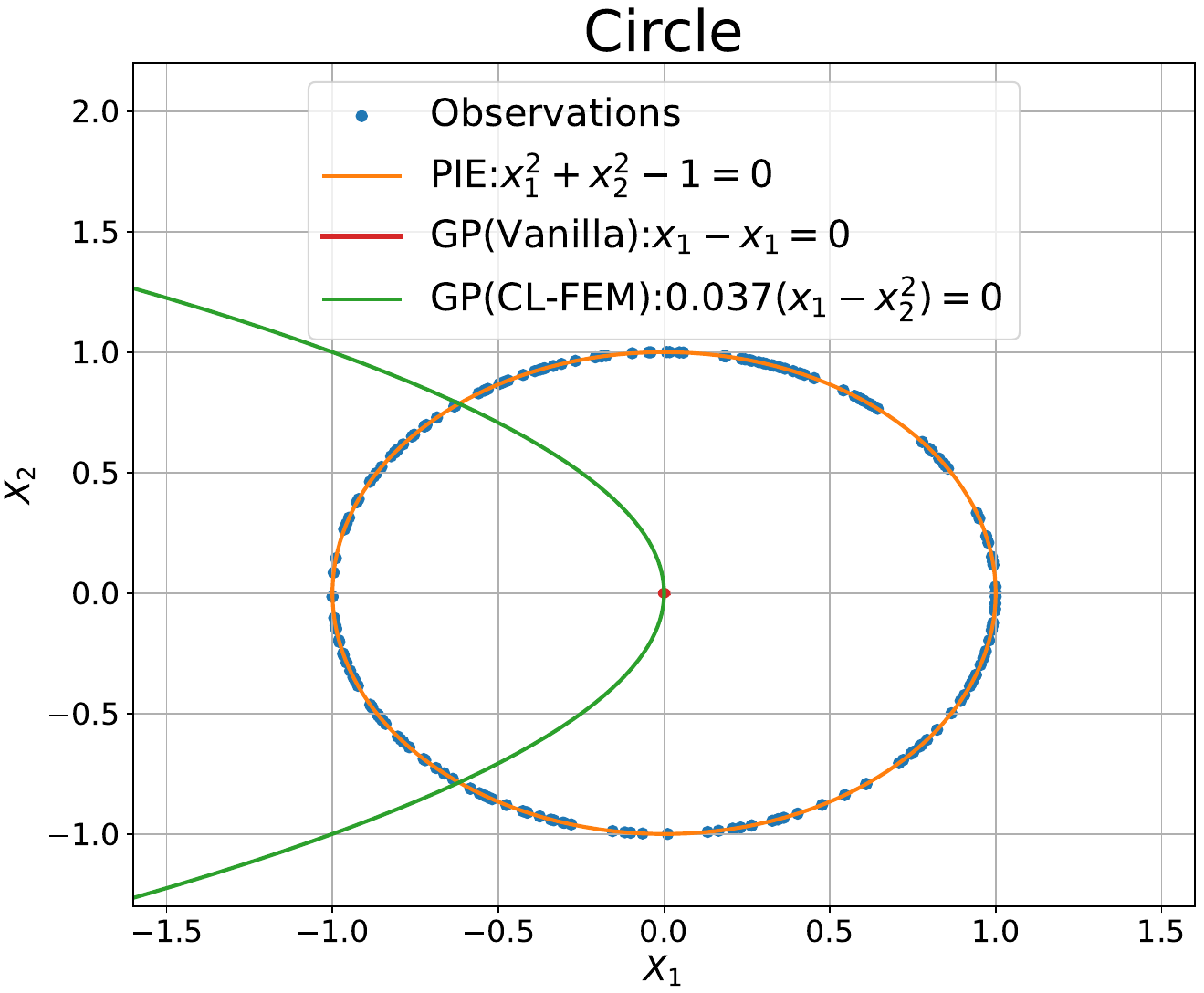}}
     \subfigure[Ideal gas law.]{
        \label{fig: example3}
        \includegraphics[width=0.28\textwidth]{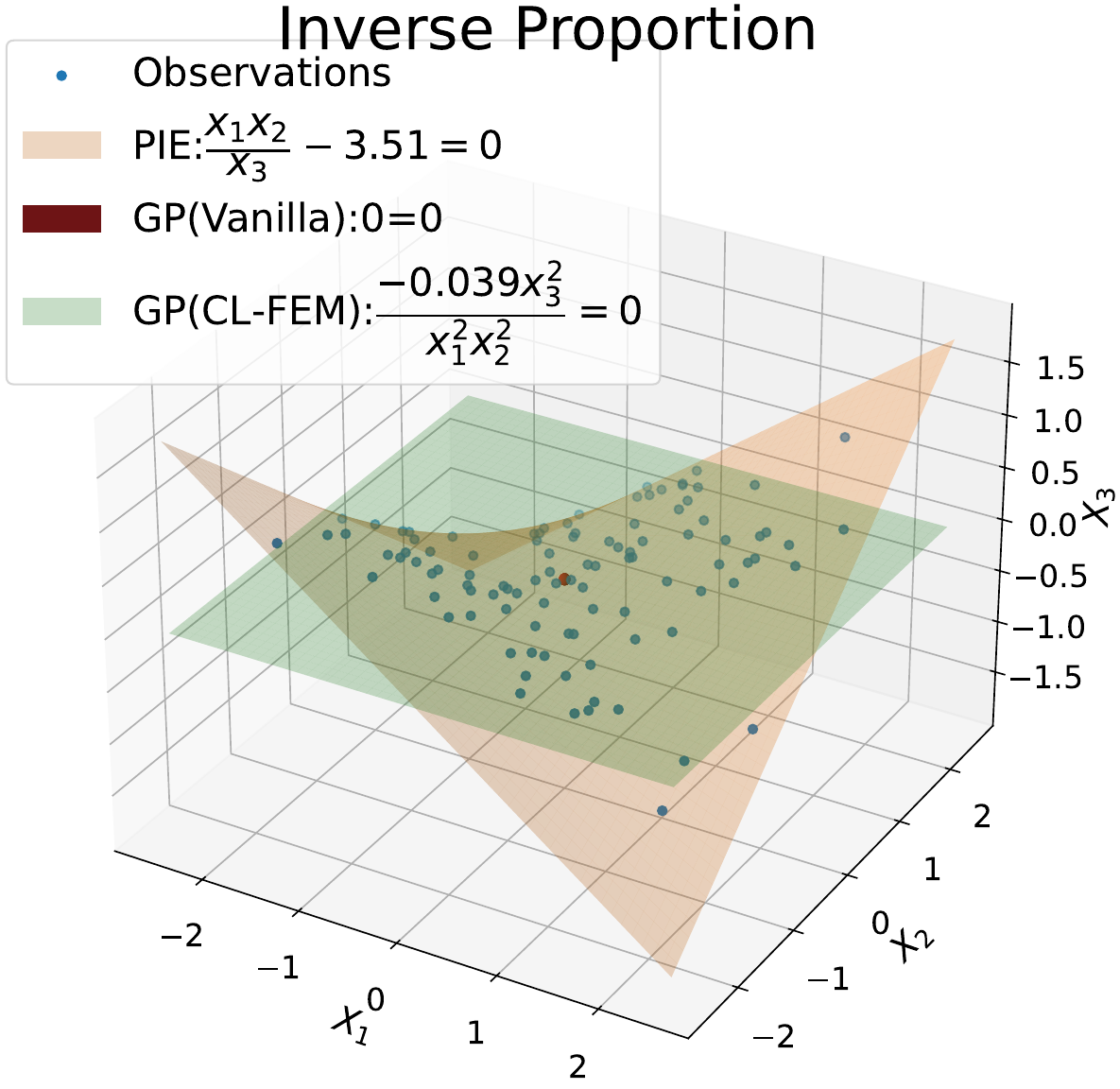}}
\vspace{-2mm}
\caption{Three scientific discovery examples on unsupervised data following specific implicit equations. Specifically, Figure \ref{fig: example1} is a hyperbola equation for particle trajectories, Figure \ref{fig: example2} is a  elliptical equation for celestial motion, and Figure \ref{fig: example3} is a inverse proportion equation. See the legend for their complete forms. Compare the learned implicit equations, we conclude that PIE \textit{effectively} tackles the  degenerate solutions and \textit{significantly} outperforms all the other baselines.}
\vspace{-5mm}
\label{fig: examples}
\end{figure}

However, due to the dense distribution of degenerate solutions, directly applying SR approaches to this task can lead to severely low performance \cite{db-fem,cl-fem,implicit-enhanced-gp}. 
Rather than learning an explicit function $y=f(\mathbf{x})$, classic SR approaches need to be slightly modified to learn an implicit equation satisfying $f(\mathbf{x})=0$ on all the input data. 
However, besides the ground-truth equation $f$, there are also a large amount of degenerate solutions in the discrete search space. 
These degenerate solutions can take a \textit{rich} set of different forms, e.g., $f(\mathbf{x})=g(\mathbf{x})-\hat{g}(\mathbf{x})$ for semantically equivalent $g$ and $\hat{g}$, $f(\mathbf{x})=C g(\mathbf{x})$ for $C=0$ and arbitrary $g$, and $f(\mathbf{x})= g(\mathbf{x})/C$ for extremely large $C$ and arbitrary $g$. 
This makes classic SR approaches easily fall into degenerate solutions. 
Several enhanced approaches are proposed to tackle this problem  \cite{db-fem,cl-fem}. 
Generally, the core idea is to revise the fitness functions to explicitly penalize degenerate solutions in different ways. 
However, for real-world applications like recovering physical equations from the AI-Feynman dataset \cite{ai-feynman}, we find they still fall into approximately degenerate solutions due to the highly discrete and ill-conditioned search space  (Figure \ref{fig: examples}). Currently, \textit{few existing approaches customized for this task achieve satisfactory performance}.

To tackle this problem, we introduce a novel pre-training framework---namely, \textbf{P}re-trained neural symbolic model for \textbf{I}mplicit \textbf{E}quation (PIE)---to predict implicit equations from given unsupervised data end-to-end. 
Intuitively, an experienced expert can easily infer what the underlying expression is given a set of visualized data points from an elliptical equation (like that in Figure \ref{fig: example2}). 
This motivates us that, instead of designing new fitness functions to penalize degenerate solutions, we can \textit{utilize the prior learned from pre-training datasets to infer the skeleton directly}. 
Thus, rather than searching in a discrete and high-dimensional equation space from scratch, we model this task as a translation task from the unsupervised data to the skeleton of the underlying symbolic equation. 
We train a Transformer model via a randomly generated large pre-training dataset for end-to-end prediction, and we combine beam search with a comprehensive learning fitness function \citep{cl-fem} to further learn non-degenerate constants. 
Extensive experiments show that, leveraging the prior from a pre-trained model, PIE \textit{effectively} tackles the problem of degenerate solutions and \textit{significantly} outperforms \textit{all} the enhanced SR approaches on both in-domain and out-of-domain datasets. 

We summarize the appealing features of PIE as follows. 
(1) The novel approach. Instead of revising the fitness function, PIE formulates implicit equation discovery as a  translation task to utilizes the learned prior to \textit{effectively} avoid degenerate solutions. 
(2) The high performance. Compared with existing SR approaches, PIE achieves \textit{significantly} higher performance in terms of the learning accuracy. 
(3) The efficiency. PIE predicts the underlying equation {end-to-end} from the input data, without time-consuming searching for the equation skeleton in the high-dimensional space. 
(4) The scalability. The high performance on both in-domain and out-of-domain datasets illustrates the high scalability of PIE. 
(5) The robustness. PIE is relatively robust to noisy data and the number of input data points. 
To the best of our knowledge, we are the \textit{first} to introduce pre-training language models for scientific discovery and achieve \textit{highly} encouraging performance  on real-world complex datasets.

\section{Preliminary}
\subsection{Implicit Function Discovery on Unsupervised Data}
An implicit function represents a mathematical relation where the dependency is not given explicitly:
\begin{align}\label{eq:implicit-fun}
    f(\mathbf{x})=0,
\end{align}
where $\mathbf{x}=(x_1,x_2,\cdots,x_D)$ are independent variables of the system. 
Notably, implicit functions in this form may not be possible to solve for \textit{any} single variable in the form of \(y=f(\mathbf{x})\). This implies that implicit functions can be more expressive and concise in defining closed surfaces or functions with multiple outputs than explicit functions.

The need to discover implicit functions arises when we have a series of data points without labeling the input and output variables. Our goal is to identify the \(f(\mathbf{x})\) that satisfies equation~\eqref{eq:implicit-fun} based on the provided unsupervised data. Here, we provide some physical examples:
\begin{enumerate}[label=(\alph*)]
    \item Particle motions: In special relativity, the particle's path through spacetime can be modeled as a hyperbola in time \(T\) and the spatial location variable \(X\). The motion can be described as \(X^2-c^2T^2=\left(\frac{mc}{F}\right)^2\), where \(m\) is the mass of the particle, \(F\) is a constant force exerted on the particle, and \(c\) is the speed of light~\cite{franklin2014fields}.
    \item Astrodynamics: In the study of astrodynamics, the motion of astronomical bodies, such as planets, moons, and comets, can be described as an elliptical orbit of the equation \(\frac{4x_1^2}{(r_1+r_2)^2} + \frac{x_2^2}{r_1r_2} = 1\), where \((x_1, x_2)\) represent the position coordinates, and \(r_1,r_2\) are the distance of the closest and furthest approaches, respectively~\cite{wilson1968kepler}.
    \item Heat conduction: The process of heat transfer in the atmosphere can be represented as implicit equations of the form \(\frac{\partial u}{\partial t} = \kappa \nabla^{2} u\), where \(u\) is the temperature, \(t\) is time, \(\kappa\) is the thermal diffusivity, and \(\nabla^{2}\) represents the Laplacian operator~\cite{gumerov2005fast}.
\end{enumerate}
Thus, implicit function discovery is of great importance in various scientific and engineering fields, as it allows researchers to uncover the underlying mathematical relationships governing complex natural phenomena, without the need for explicitly labeled or supervised data. By recovering the implicit equations from unsupervised data, scientists can gain deeper insights into the fundamental principles governing the systems under investigation, leading to improved predictive models, enhanced understanding of physical processes, and more accurate simulations.

\subsection{Symbolic regression}
Symbolic regression (SR) discovers a symbolic equation and tunes its corresponding parameters to model an experimentally obtained dataset. Formally, given a dataset $\mathcal{D}=\{\mathbf{x}_n\}_{n=1}^{N_\mathcal{D}}$,
SR for implicit functions aims to identify a function $f$ such that $f(\mathbf{x}_n)=0,\;\forall n= 1,\ldots,N_\mathcal{D}$.
Unlike neural networks with numerous parameters and activation functions, a symbolic equation consists of expressions with higher interpretability, e.g. \(f(x)=x-\sin(x)\). This property facilitates understanding and makes SR more suitable for tasks such as inferring physical laws from data.
Specifically, a symbolic equation comprises three components: a pre-defined list of operators such as $(+, -, \log, \sin, \ldots)$, input variables $(x_1, x_2, \ldots)$, and numeric constants like $(1.414, 2.735, \ldots)$. 
In our paper, we perform SR in two phases. First, we predict the symbolic skeleton $\tilde{f}$ of a target function $f$, where the skeleton $\tilde{f}$ is defined by replacing numeric constants in $f$ with the constant placeholder $\diamond$. For instance, the skeleton of function $f(x)=3.132x+\sin(x-2.173)$ is $f(x)=\diamond x+\sin(x-\diamond)$. Second, we estimate the numeric values of the constants in the skeleton using a typical optimization technique, namely the Broyden–Fletcher–Goldfarb–Shanno algorithm (BFGS).

\subsection{Set Transformer} \label{sec: set-transformer}
The Set Transformer \citep{set-transformer} is designed for permutation-invariant function approximation. It is built upon the attention mechanism and is capable of capturing complex interactions within unordered sets of data. The Set Transformer consists of four key components:
\begin{itemize}[leftmargin=0.5cm]
    \item Multihead Attention Block (MAB): the core building block of the Set Transformer, which computes attention-based interactions between two input sets \(X\) and \(Y\):
   \begin{align}
       \text{MAB}(X,Y) =\text{LayerNorm}(H +\text{rFF}(H)),
   \end{align}
   where $H=\text{LayerNorm}(X +\text{Multihead}(X, Y, Y; \omega))$, $\text{rFF}$ is a row-wise feedforward layer, and $\omega$ is the activation function used for the computation of attention weights.
   \item Set Attention Block (SAB):
   it applies the MAB to the input set \(X\) in a self-attention manner, allowing the model to capture pairwise interactions among elements in the set and encode a set of features with equal size.
   The SAB can be expressed as:
   \begin{align}
       \text{SAB}(X) = \text{MAB}(X, X).
   \end{align}
     \item Induced Set Attention Block (ISAB): it reduces the computational complexity compared to the SAB and introduces a set of learnable inducing points that serve as a compressed representation of the input set. The ISAB can be expressed as:
     \begin{align}
         \text{ISAB}(X) = \text{MAB}(X, \text{MAB}(I,X)),
     \end{align}
     where $I$ are the inducing points.
     \item  Pooling by Multihead Attention (PMA): it applies the MAB to the feature set \(Z\) constructed from an encoder and a learnable set of \(k\) seed vectors \(S\), serving as an aggregation scheme for the entire permutation-invariant network.
     The PMA can be expressed as:
     \begin{align}
         \text{PMA}_k(Z) = \text{MAB}(S, \text{rFF}(Z)).
     \end{align}
\end{itemize}

The Set Transformer is a universal function approximator in the space of permutation-invariant functions, as shown in the paper~\cite{set-transformer}. This property, combined with its ability to capture complex interactions within sets, makes the Set Transformer a powerful tool for a variety of tasks involving unordered data, such as point cloud processing, graph neural networks, etc.

\section{Method} \label{sec: method}

\begin{figure}[t]
\centering
\includegraphics[width=0.88\textwidth]{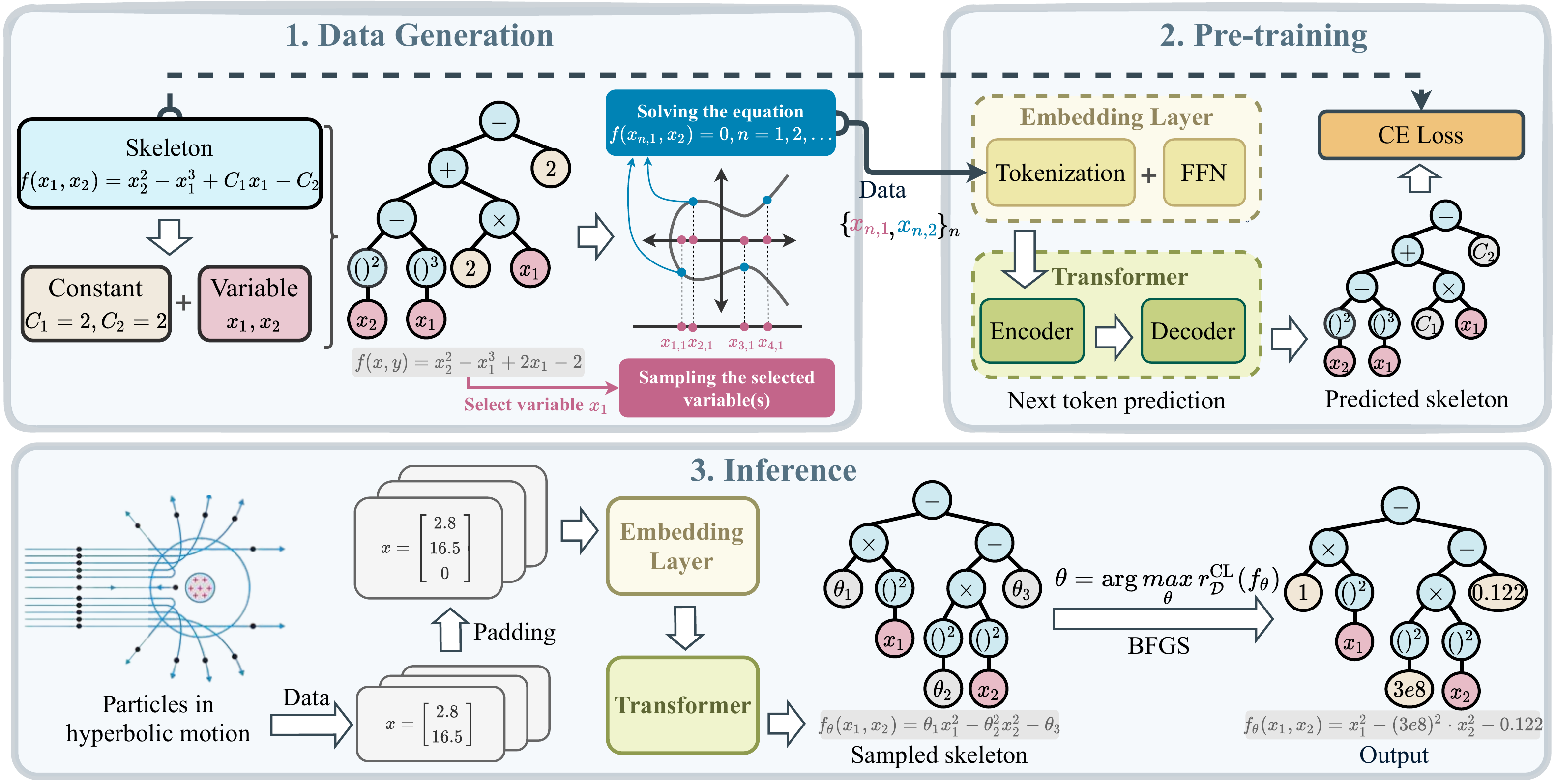}
\caption{
Illustration of the framework of PIE for symbolic implicit equation discovery. 
Part 1 shows the generation process of the pre-training dataset $\mathcal{D}_{\text{pre-train}}=\{(\mathcal{D}_k, \tilde{f}_k)\}_{k=1}^{K}$; 
Part 2 shows the model architecture, which contains an embedding layer to handle raw floating-point data \citep{nesymbres} and a Set Transformer \citep{set-transformer} to translate the input to the skeleton; 
Part 3 shows the inference process, which combines the beam search with the CL-FEM fitness \citep{cl-fem} to further avoid degenerate solutions.
}
\label{fig: visualization}
\vspace{-6mm}
\end{figure}

Implicit equation discovery aims to recover mathematical relation from given unsupervised data. 
However, most existing approaches suffer from the  highly discrete and ill-conditioned search space caused by degenerate solutions. 
In this section, we introduce a novel pre-training framework, namely, \textbf{P}re-trained neural symbolic model for \textbf{I}mplicit \textbf{E}quation (PIE). 
The core idea of PIE is to learn a prior from a large pre-training dataset to avoid degenerate solutions. We illustrate our framework in Figure \ref{fig: visualization} and provide the pre-training and inference pseudo-code in Algorithm~\ref{alg:pie-pre-training} and~\ref{alg:pie-inference} in Appendix.

\subsection{Problem Setup}

\textbf{Modifying SR for Implicit Equation Discovery}
Intuitively, SR is a natural tool to tackle this task. We can simply modify classic SR approaches to learn an implicit equation satisfying $f(\mathbf{x})=0$ rather than the original target $y=f(\mathbf{x})$ on data points $\mathcal{D}=\{\mathbf{x}_n\}_{n=1}^{N_\mathcal{D}}$. Then, the fitness can be defined as:
\begin{equation} \label{eq: fitness-vanilla}
    r_\mathcal{D}(f) = -\frac{1}{N_\mathcal{D}}\sum_{n=1}^{N_\mathcal{D}}\|f(\mathbf{x}_n)\|_\ell,
\end{equation}
where the $\ell$-norm is typically used for $\ell=1,2$. 
Then, SR approaches can search for the implicit equation in the discrete equation space to select one that maximizes $r_\mathcal{D}$. 

\begin{wrapfigure}{r}{0.4\textwidth}
    \vspace{-3mm}
    \centering
    \includegraphics[width=0.4\textwidth]{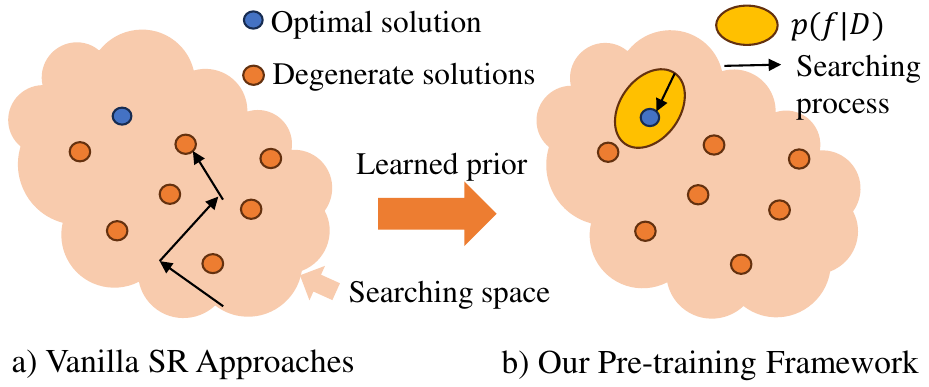}
    \vspace{-4mm}
    \caption{Visualize the limitation of vanilla SR approaches and the intuition of our pre-training framework. 
    }
    \label{fig: intuition}
    \vspace{-3mm}
\end{wrapfigure}

\textbf{Limitation} 
Besides the ground-truth equation, there are also a large amount of degenerate solutions in the discrete search space that satisfy the optimal condition. We list several frequently occurring degenerate solutions in practice. 
1) $f(\mathbf{x})=g(\mathbf{x})-\hat{g}(\mathbf{x})$ for all the semantically equivalent $g$ and $\hat{g}$, e.g., $g(\mathbf{x})=\sin^2(x_i) + \cos^2(x_i)$ and $\hat{g}(\mathbf{x})=1$, and $g(\mathbf{x})=C_1x_i + C_2x_j$ and $\hat{g}(\mathbf{x})=x_j + x_i/C_3$ (easily to be optimized to $C_1=C_2=1/C_3\neq 0$). 
2) $f(\mathbf{x})=Cg(\mathbf{x})$ for $C=0$ and arbitrary $g$. 
3) $f(\mathbf{x})=g(\mathbf{x})/C$ for extremely large $C$ and arbitrary $g$. 
They are dense distributed in the search space and usually take different forms. Such highly discrete and ill-conditioned search space makes SR approaches extremely easily to achieve severely low performance. 
Several enhanced approaches are proposed in previous research via revising the vanilla fitness functions to explicitly penalize degenerate solutions (see Appendix \ref{app: related-work}). 
However, due to the highly discrete and ill-conditioned search space, we find these approaches still  fall into approximately degenerate solutions (see Figure \ref{fig: examples} and Table \ref{tab: in-domain}). 
Thus, {few existing approaches customized for this task achieve satisfactory performance currently}.

\textbf{Intuition of PIE} 
Given a set of visualized data points from an elliptical equation (like that in Figure \ref{fig: example2}), an experienced expert can easily infer what the underlying expression it follows, and then he can directly use numerical optimization tools to fit the constants of the inferred skeleton. 
Intuitively, the former can be viewed as a translation process. That is, the expert translates the visualized data points to some elliptical equation based on his prior knowledge, and degenerate solutions will not be preferred by him as these contradict his intuition and experience. 
This motivates us that, to avoid SR searching in the highly discrete and ill-conditioned space, we can pre-train a ``language model'' for such ``translation'' task to predict the skeleton of the underlying implicit equation and then use inner optimization  to further determine the constants (see Figure \ref{fig: intuition} for visualized intuition). 

\subsection{Pre-Training Data Generation} \label{sec: data}

For the pre-training dataset, each training sample is comprised of a ``skeleton'' $\tilde{f}_k$ of an implicit equation $f_k$ as the target, and a corresponding set of data points $\mathcal{D}_k=\{\mathbf{x}_n\}_{n=1}^{N_{\mathcal{D}_k}}$ as the input. 
Each data point in $\mathcal{D}_k$ satisfies the condition $f_k(\mathbf{x}_n)=0$, for $n=1,2,\cdots,N_{\mathcal{D}_k}$. 
The generation of these training samples involves two phases. At phase one, a random equation $f$ is sampled. 
At phase two, a set of data points are sampled by solving the equation $f(\mathbf{x})=0$. 
The skeleton $\tilde{f}_k$ of equation $f$ along with the sampled dataset $\mathcal{D}_k$ then form the training sample, i.e., $\mathcal{D}_{\text{pre-train}}=\{(\mathcal{D}_k, \tilde{f}_k)\}_{k=1}^{K}$. 

\textbf{Sampling equations.} 
The sampling process of equation $f$ is guided by the framework presented by \citet{lample2019deep}. 
\textit{First}, we generate implicit equations $f$ using the representation of symbolic trees with a depth first traversal. That is, the symbolic trees we generated are represented by sequences of symbols in a prefix notation \cite{nesymbres}. 
These randomly generated trees consist of non-leaf nodes representing mathematical operators and leaf nodes representing constants or variables. 
We report all the mathematical operators we use and their corresponding (unnormalized) probabilities to be chosen during the symbolic tree generation process in Table \ref{tab: operator_probs}. 
If a constant is selected at the leaf node, we sample its value from a standard normal distribution. 
\textit{Then}, we modify the sampled equations $f$ to their corresponding ``skeleton'', denoted as $\tilde{f}$, in which all the  numerical constants are substituted with placeholders $\diamond$. An example implicit equation and its skeleton are $f(\mathbf{x})=0.5x_1+3.0x_2$ and $\tilde{f}(\mathbf{x})=\diamond x_1+\diamond x_2$, respectively. Note that we will use $f$ to generate data points and $\tilde{f}$ as the prediction target in the pre-training dataset.

\textbf{Sampling Data Points} For each implicit equation $f_i$ sampled above, there are three steps to  sample the data points. \textit{First}, we uniformly select one dimension $i$ from the input vector $\mathbf{x}$ of $f(\mathbf{x})$. \textit{Then}, for all the other dimensions $j$ that satisfies $j\neq i$, we sample their values from a normal distribution $\mathcal{N}(0, 1)$. \textit{Finally}, for $x_i$, we determine it by solving the equation $f(\mathbf{x})=0$.
For each equation $f$, we generate a set of $N$ data points repeatedly. 
Noting that a solution for the equation $f(\mathbf{x})=0$ may not always exist. In cases where $N$ input points cannot be sampled within a specified number of trials, we regard the equation $f$ as ill-conditioned and directly discard it. 
We further pad the input vectors $\mathbf{x}$ whose dimensions are smaller than the maximal possible dimension $D_{\max}$ with zeros to yield fixed-dimensional inputs.
See more details about the sampling process in Table \ref{tab:hyperparameters} in Appendix. 

\subsection{Model architecture}  

Specifically, the model architecture employed in PIE is based on Set Transformer (as described in Section \ref{sec: set-transformer}), and it consists of three main components. That is, an embedding layer to handle raw floating-point data, a set encoder to encode the processed data points, and a Transformer decoder to generate expressions. We report all the hyperparameters used in our model architecture in Table \ref{tab:hyperparameters}.

\textbf{The Embedding Layer} 
Similar to that in \citet{nesymbres},  the embedding layer converts raw floating-point data points to learned embeddings. \textit{First}, it converts each variable's value from a floating-point representation to a multi-hot bit representation of dimension $d_{\text{emb}}$ based on the half-precision IEEE-754 standard \citep{ieee754}. Thus, each data point is converted to a vector with dimension $D_{\max} \times d_{\text{emb}}$.
\textit{Then}, it employs a two-layer multilayer perceptron (MLP), which further projects the vectors above to learned embeddings with dimension $d_{\text{hid}}$. The embeddings are then fed into the Transformer below. 

\textbf{The Encoder and Decoder} The encoder and decoder of PIE are based on the Set Transformer and the regular Transformer, respectively \citep{set-transformer,nesymbres}. 
We use the official implementation of the encoder  provided by \citet{set-transformer}. 
Specifically, the encoder maps the set of embeddings mentioned above into latent vectors, and the decoder autoregressively generates the prefix notation $\bar{f}$ based on the outputs of the encoder. 
During training, the objective of PIE is to minimize the cross-entropy (CE) loss between the ground-truth skeleton $\tilde{f}$ and predicted one $\bar{f}$. 
See more details about the training in Appendix \ref{app: implementation}.

\subsection{Inference}

During inference, the decoder computes a probability distribution $p(\bar{f}_{l+1} \mid \bar{f}_{1:l}, \mathbf{z})$  and selects the next symbol from this distribution iteratively. 
Similar to that in natural language processing tasks, we employ beam search with size $N_{\text{BS}}$ to further improve the inference accuracy. 
After obtaining $N_{\text{BS}}$ skeleton candidates, we need to select the best one by recovering the skeletons to complete equations and comparing which fits the input data best. We use the BFGS algorithm \citep{bfgs} for inner optimization to recover the constants by maximizing the fitness function.

However, we note that even with a precise skeleton predicted by the pre-trained model, directly optimizing the objective in Equation (\ref{eq: fitness-vanilla}) can still lead to degenerate solutions. For example, given a fully correct skeleton of the hyperbola equation $\bar{f}=C_1 x_1^2 - C_2 x_2^2 - C_3=0$, directly optimizing Equation (\ref{eq: fitness-vanilla}) can still lead to degenerate solutions like $C_1=C_2=C_3=0$. Thus, to further avoid degenerate solutions, we employ the comprehensive learning fitness evaluation mechanism (CL-FEM) introduced in \citep{cl-fem} for constant optimization. 

\textbf{CL-FEM for Constant Optimization} The core idea of CL-FEM is to explicitly penalize degenerate solutions at each dimension via the fitness function \citep{cl-fem}. 
Specifically, calculating CL-FEM fitness involves three steps. \textit{First}, a stochastic dataset is generated for each \textit{valid} dimension (i.e., non-padding dimension). For the $j^{\text{th}}$ dimension, we create the stochastic dataset $\mathcal{D}^j=\{\mathbf{x}_n^j\}_{n=1}^N$ via:
\begin{equation}
x_{n,i}^j = 
\begin{cases} 
\text{rand}(L, U), & \text{if } i = j; \\
x_{n,i}, & \text{otherwise.}
\end{cases}
\end{equation}
Here, $\text{rand}(L, U)$ is a function that generates a random number within the range $(L, U)$. We use $L=-1$ and $U=1$, respectively. The term $x_{n,i}^j$ denotes the $i^{\text{th}}$ dimension of the vector $\mathbf{x}_n^j$, and $x_{n,i}$ is the $i^{\text{th}}$ dimension of $\mathbf{x}_n$ from the original dataset. 
The stochastic $\mathcal{D}^j$ is used to test whether an implicit equation is degenerate at $j^{\text{th}}$ dimension. 
\textit{Second}, for each dimension $j = 1, 2, \cdots, D$, we calculate the ``distance'' between the real dataset $\mathcal{D}$ and the stochastic dataset $\mathcal{D}^j$ via:
\begin{equation}
d_{j} = \frac{1}{N} \sum_{n=1}^{N} (\bar{f}(\mathbf{x}_n) - \bar{f}(\mathbf{x}_n^j))^2.
\end{equation}
As $\mathbf{x}_n$ differs from $\mathbf{x}_n^j$ only at the $j^{\text{th}}$ dimension, a non-degenerate function $\bar{f}$ at dimension $j^\text{th}$ is indicated by $d_{j} > 0$. 
In practice, a small positive threshold $\tau$ is employed as a tolerance level \citep{cl-fem}. 
\textit{Finally},  the CL-FEM fitness function is defined as:
\begin{equation}
r_\mathcal{D}^\text{CL}(f) =
\begin{cases}
-\infty, & \text{if } d_{j} \le \tau \text{ for some } j; \\
-\frac{1}{N_\mathcal{D}}\sum_{n=1}^{N_\mathcal{D}} \|f(\mathbf{x}_n)\|_\ell, & \text{otherwise}.
\end{cases} \label{eq: cl-fem}
\end{equation}
We set $\tau$ and $\ell$ as that in \citet{cl-fem} and report all the hyperparameters in Table \ref{tab:hyperparameters} in Appendix, and we conduct ablation to compare our approached equipped with CL-FEM (PIE) to that equipped with Equation (\ref{eq: fitness-vanilla}) (PIE(Vanilla)) to show its effectiveness for constant optimization in Table \ref{tab: in-domain}.

\section{Experiments} \label{sec: experiments}
We conduct extensive experiments\footnote{We will all the codes of our approach and the pre-trained model once the paper is accepted.} on our framework to 1) evaluate the in-domain performance on a synthetic dataset generated from the same distribution as the pre-training dataset; 2) evaluate the out-of-domain generalization ability on equations revised from the AI-Feynman database \citep{ai-feynman}; 3) evaluate its robustness under data with different noise intensity and input sizes; 4) evaluate the sensitivity of different hyperparameters during the training and the inference.

\subsection{Experimental Setups}
\textbf{Dataset} 
We evaluate the performance of our model on several datasets. The first dataset, named as the Synthetic dataset, tests the in-domain performance of our model. This dataset is created via the same process as the dataset we used to pre-train our model. The second dataset is the AI-Feynman dataset, which is derived from the well-known Feynman SR database \cite{ai-feynman}, a real-world dataset widely used to evaluate the performance of symbolic regression models in previous research \citep{nesymbres,symb}. We adapt the original equations from this database into implicit equations. Specifically, for each equation $f(\mathbf{x})$ from the origin database, we sample a positive constant $c$, subtract it from the equation, and formulate the implicit equation $f(\mathbf{x}) - c = 0$. The constant is used to avoid situations where $f(\mathbf{x})=0$ yields no solutions, such as $\exp(-\frac{1}{2}(x_2-x_1)^2)=0$. Finally, we exclude equations involving more than three variables to align with the maximum number of variables our model can accommodate. This setting is also widely employed in previous research like \citet{nesymbres}. The detailed description of equations in our datasets is available in Appendix \ref{app: benchmark}.

\textbf{Baselines} We compare our model against four different baselines. We adopt baselines from genetic programming (GP) and deep symbolic optimization (DSO) for implicit equation discovery, which represent classical SR approaches in GP-based and RL-based paradigms, respectively. 
For each approach, we employ two fitness functions, i.e., the Vanilla fitness function in Equation \ref{eq: fitness-vanilla} and the CL-FEM fitness function in Equation \ref{eq: cl-fem} (the SOTA previous approach). 
Thus, we obtain four baselines in total to represent different classic SR approaches. See Appendix \ref{app: baseline} for implementation details.

\textbf{Metric} 
Directly comparing the ground-truth $f$ and the learned $\bar{f}$ is intractable, as equations in completely different forms can be semantically equivalent. For example, only considering the arrangement order of $\mathbf{x}$, $f(\mathbf{x})=x_1+x_2+x_3$ can have six different forms that are semantically equivalent. 
Previous research on implicit equation discovery typically uses the fitness functions (vanilla or enhanced ones) as the metric for evaluation. However, these fitness functions have certain limitations (e.g., Vanilla favors degenerate solutions and CL-FEM favors approximately degenerate solutions), which prevents them from fair evaluation. Thus, we propose a \textit{novel} metric to measure the similarity between the learned $\hat{f}$ and the ground-truth $f$. This metric is based on a simple but effective intuition: points sampled from a better $\hat{f}(\mathbf{x})=0$ should be ``closer'' to the hyperplane of $f(\mathbf{x})=0$. Thus, we define the metric between $f$ and $\bar{f}$ as:
\begin{equation}
    \label{mse}
    MSE(\hat{f}, f)=\mathbb{E}_{\mathbf{x}\sim \hat{f}(\mathbf{x})=0}|f(\mathbf{x})|^2,
\end{equation}
Since $f(\mathbf{x})=0$ and $cf(\mathbf{x})=0$ are semantically equivalent implicit equations, we normalize the above MSE by dividing the ``average'' value of $f$, i.e.,
\begin{equation}
    \label{nmse}
    NMSE(\hat{f}, f)=\frac{MSE(\hat{f}, f)}{\mathbb{E}_{\mathbf{x}\sim \mathcal{N}(\mathbf{x},\mathcal{I})}|f(\mathbf{x})|^2}.
\end{equation}
Here, the NMSE ranges from $[0,\infty]$, and we further normalize it to derive our fitness metric:
\begin{equation}
    fitness(\hat{f}, f) = \frac{1}{1+\sqrt{NMSE(\hat{f}, f)}}.
\end{equation}
We further propose another metric to measure the accuracy of the learned implicit equation, i.e.,
\begin{equation}
    Acc_{\tau}=\mathds{1}(fitness(\hat{f}, f)\ge\tau).
\end{equation}
where $\mathds{1}(\cdot)$ is the indicator function, and $\tau$ is the threshold used to determine the tolerance of accuracy. 
We use $fitness$ and $Acc_{\tau}$ as our metrics to evaluate the performance of different approaches. The detailed implementation of these metrics is available in Appendix \ref{app: implementation}.

\subsection{Comparative Evaluation} \label{sec: comparative}

\begin{table}[t]
\centering
\caption{Comparative analysis of our framework against all the other baselines on the synthetic dataset. We employ multiple metrics for a comprehensive evaluation. We observe directly employing CL-FEM in SR baselines achieves severely low fitness in our metrics, especially on DSO (we discuss the potential reasons for that in the footnotes below). Furthermore, results show that: 1) PIE achieves significantly higher performance over all the SR-based baselines; 2) both the pre-trained Transformer model and the CL-FEM fitness function contribute to the high performance of PIE.}
\vspace{-2mm}
\label{Synthetic}
\begin{tabular}{@{}lcccccc@{}}
\toprule
Method      & fitness   &$\text{Acc}_{0.5}$ &$\text{Acc}_{0.7}$   & $\text{Acc}_{0.8}$ & $\text{Acc}_{0.9}$ & $\text{Acc}_{0.99}$           \\ \midrule
GP (Vanilla)         & $0.237$   &$15.0\%$   &$2.5\%$   &$2.5\%$   &$0.0\%$   & $0.0\%$  \\
GP (CL-FEM)       & $0.165$ &$15.0\%$   &$3.8\%$ &$2.5\%$  & $1.3\%$     & $0.0\%$    \\
DSO (Vanilla)        & $0.313$   &$33.8\%$   &$2.5\%$   &$0.0\%$  & $0.0\%$     & $0.0\%$   \\
DSO (CL-FEM)              & $0.077$ &$8.8\%$    &$2.5\%$    &$1.3\%$ &$0.0\%$  & $0.0\%$     \\
PIE (Vanilla)     & $0.592$  &$63.8\%$   &$48.8\%$   &$46.3\%$  & $43.8\%$     & $40.0\%$    \\
PIE          & $\textbf{0.662}$    &$\textbf{71.3\%}$  &$\textbf{56.3\%}$   &$\textbf{53.8\%}$  & $\textbf{53.8\%}$     & $\textbf{45.0\%}$       \\
 \bottomrule
\end{tabular}
\label{tab: in-domain}
\vspace{-2mm}
\end{table}

\begin{table}[t]
\centering
\caption{Comparison between PIE and  baselines on the AI-Feynman Dataset. Results show that PIE consistently achieves SOTA performance by a significant margin on the out-of-domain dataset.}
\vspace{-2mm}
\begin{tabular}{@{}lccccccc@{}}
\toprule
Method      & fitness   &$\text{Acc}_{0.5}$ &$\text{Acc}_{0.7}$ &$\text{Acc}_{0.8}$     & $\text{Acc}_{0.9}$ & $\text{Acc}_{0.99}$            \\ \midrule
GP (Vanilla)         & $0.308$   &$15.4\%$   &$2.6\%$   &$0.0\%$  &   $0.0\%$   & $0.0\%$  \\
GP (CL-FEM)       & $0.138$ &$12.8\%$   &$2.6\%$    &$2.6\%$  & $0.0\%$     & $0.0\%$   \\
DSO (Vanilla)        & $0.251$   &$15.4\%$   &$0.0\%$    &$0.0\%$  & $0.0\%$     & $0.0\%$    \\
DSO (CL-FEM)              & $0.072$ &$2.6\%$    &$0.0\%$    &$0.0\%$  & $0.0\%$     & $0.0\%$   \\
PIE (Vanilla)     & $0.737$   &$69.2\%$  &$69.2\%$  &$69.2\%$  & $66.7\%$     & $61.5\%$    \\
PIE           & $\textbf{0.784}$    &$\textbf{79.5\%}$  &$\textbf{74.4\%}$  &$\textbf{74.4\%}$  & $\textbf{74.4\%}$     & $\textbf{71.8\%}$    \\
 \bottomrule
\end{tabular}
\label{tab: out-of-domain}
\vspace{-5mm}
\end{table}

\textbf{In-domain Performance }We evaluate the in-domain performance of our proposed PIE model on the Synthetic dataset and compare the results with the baselines. As shown in Table \ref{tab: in-domain}, the PIE framework, when equipped with both Vanilla and CL-FEM fitness functions, demonstrates a \textit{significant} superiority over all the baselines in terms of different metrics. Notably, all the existing SR approaches (GP and DSO) hardly ever succeed in uncovering implicit equations within this dataset
\footnote{Directly employing CL-FEM in SR baselines achieves severely low fitness. The potential reason is that, an incorrect but non-degenerate solution tends to get a higher MSE in Equation (\ref{mse}) than a simple degenerate one.
Points sampled from degenerated solutions usually follow just a normal distribution at all dimensions, while points sampled from incorrect and non-degenerate solutions can have more extreme values at specific dimension.}
\footnote{DSO with CL-FEM suffers more severe performance decrease than GP. A potential reason is that, the CL-FEM fitness makes the landscape of the optimization objective highly non-convex, which is more intractable for gradient-based optimization approaches than zero-order optimization approaches.}. 
In contrast, our PIE framework consistently identifies a large part of these equations. Furthermore, comparisons within our model variants reveal that, both the pre-trained Transformer model and the CL-FEM fitness function effectively improve the performance of PIE.

\textbf{Out-of-domain Generalization } To evaluate the out-of-domain generalization ability of our proposed model, we conduct experiments on the AI-Feynman dataset, a collection derived from the equations presented in the \textit{Feynman Lectures on Physics} textbook. This dataset is valuable for assessing out-of-domain generalization because it comprises real-world physical equations rather than synthetically generated ones. As shown in Table \ref{tab: out-of-domain}, our PIE framework \textit{consistently} outperforms all the competing baselines by a significant margin. Interestingly, the PIE framework performs slightly better on this dataset, even though it is out-of-domain. This might be attributed to the relatively simpler equations in the AI-Feynman dataset, which are on average $23\%$ shorter than those in the Synthetic dataset.

\begin{figure}[t]
    \centering
    \subfigure{
        \label{fig: robust-noise-1}
    \includegraphics[width=0.23\textwidth]{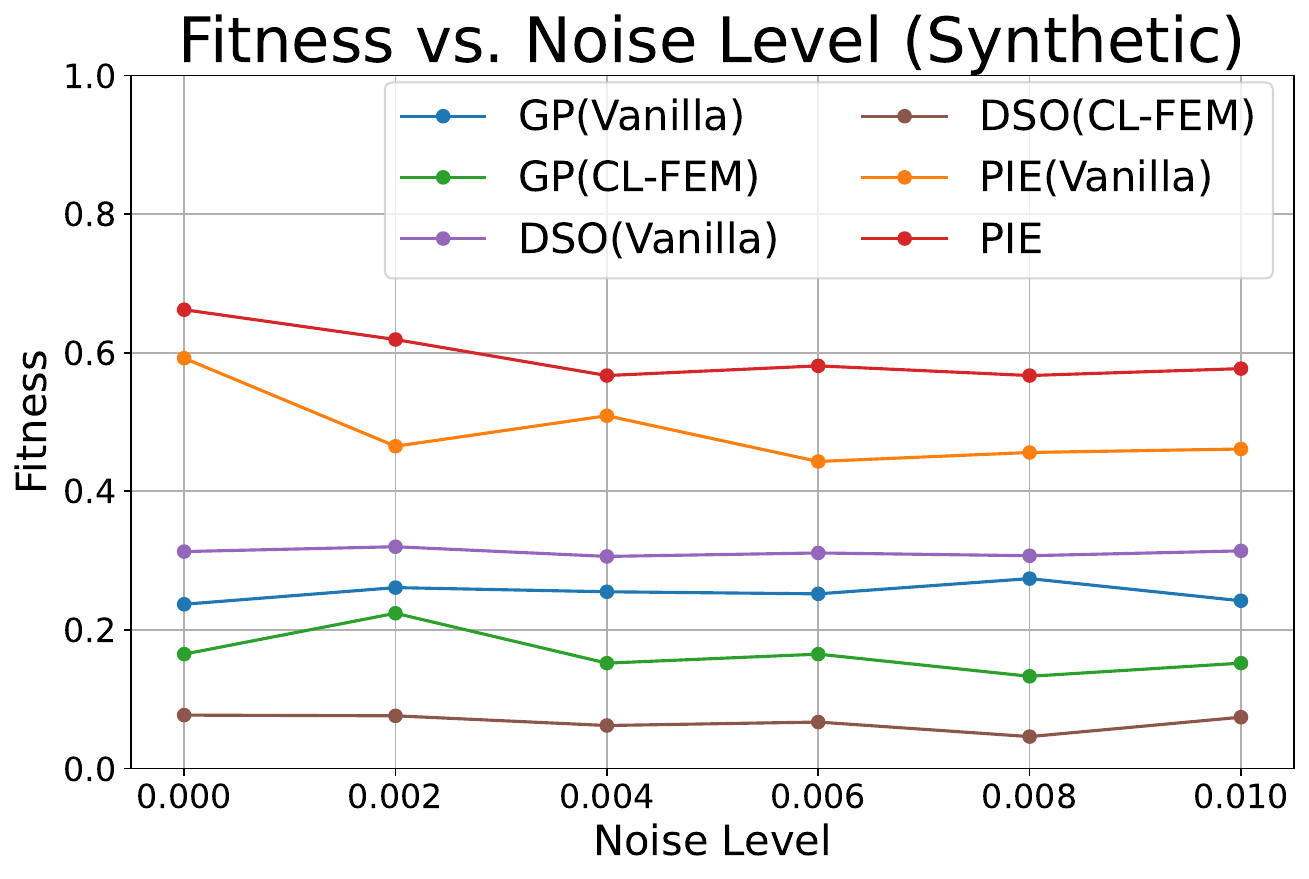}}
    \subfigure{
        \label{fig: robust-noise-2}
        \includegraphics[width=0.23\textwidth]{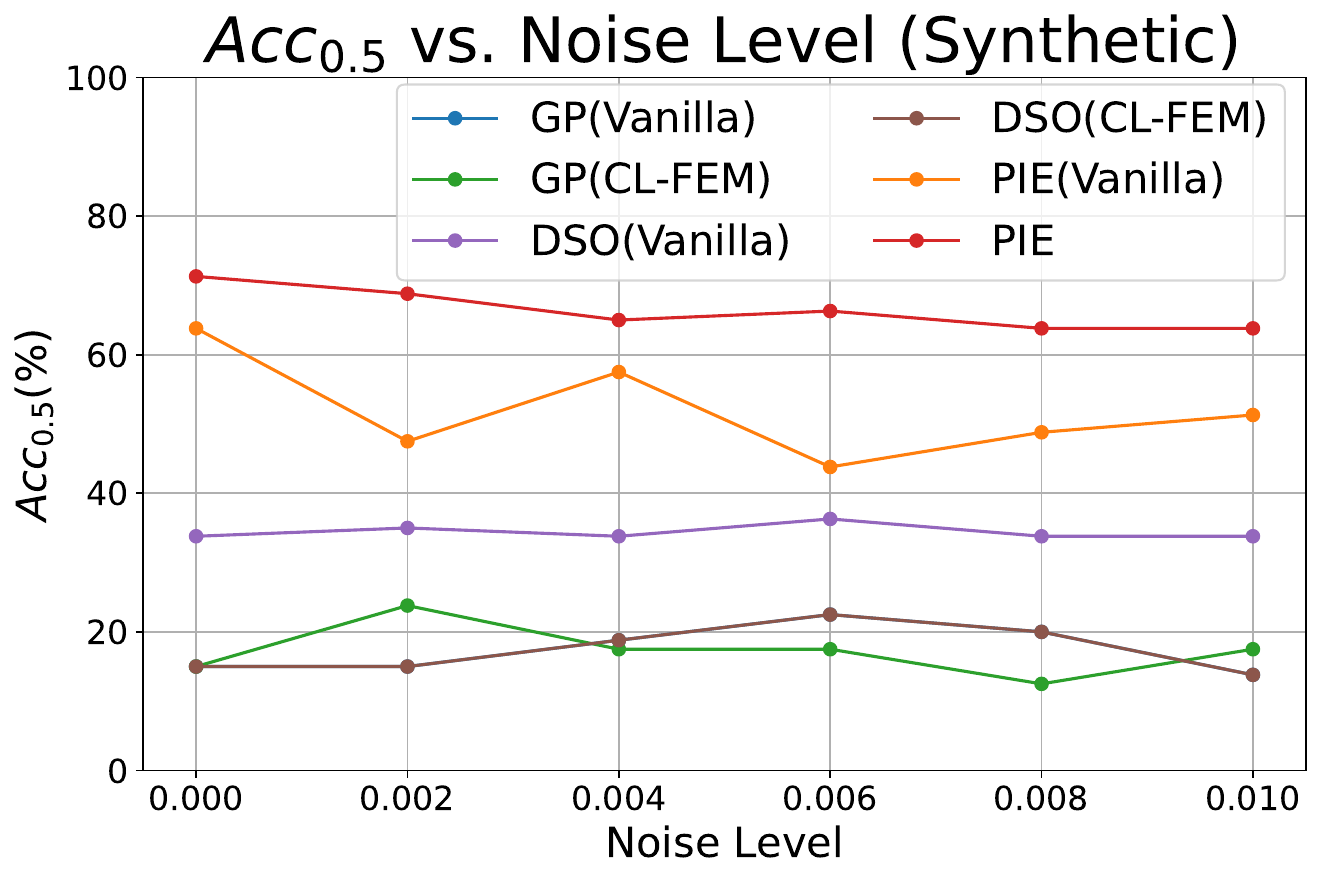}}
     \subfigure{
        \label{fig: robust-data-points-1}
        \includegraphics[width=0.23\textwidth]{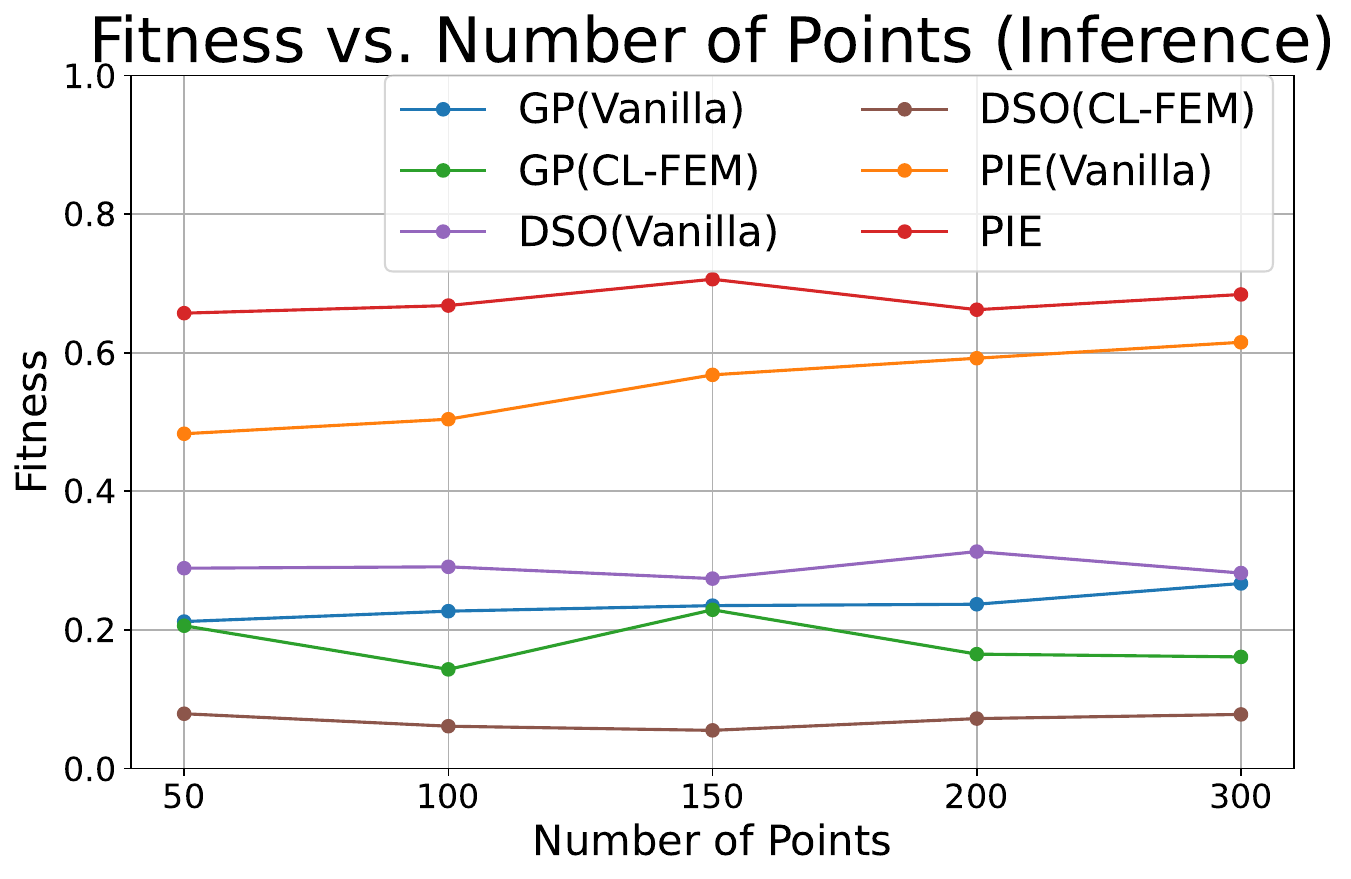}}
     \subfigure{
        \label{fig: robust-data-points-2}
        \includegraphics[width=0.23\textwidth]{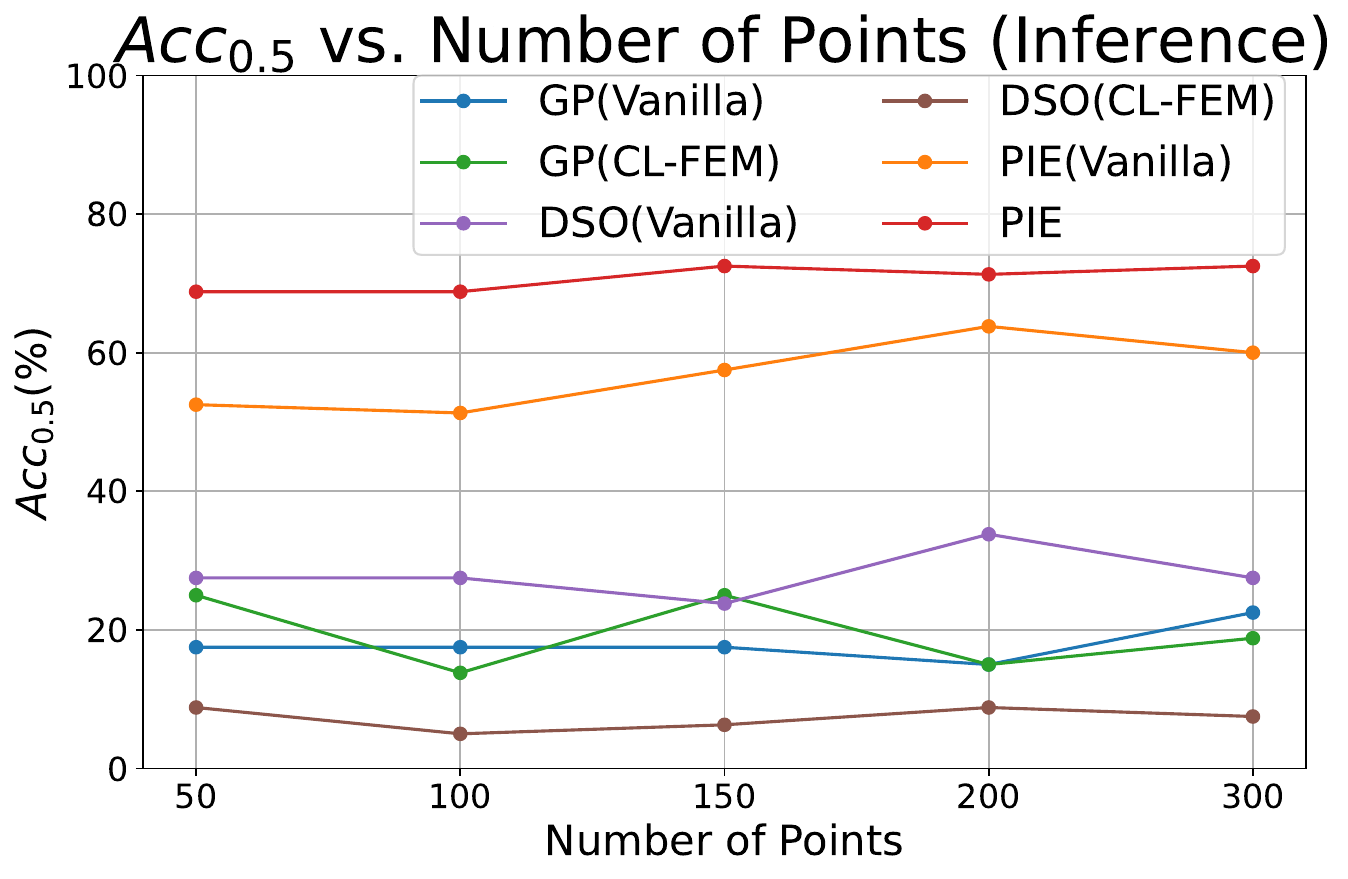}}
\vspace{-3mm}
\caption{Evaluate the robustness under noisy data and different number of data points on Synthetic. Results show that PIE is \textit{highly} robust under different noise level and number of input data points.}
\vspace{-5mm}
\label{fig: noise}
\end{figure}


\textbf{Robustness under Noisy Data} We evaluate the robustness of our proposed method in the presence of noisy data by introducing independent  Gaussian noise to each data point. Specifically, noise $\mathbf{\zeta}$  is added to each data point $\mathbf{x}$ via the transformation  $\mathbf{x} \rightarrow \mathbf{x}(1 + \mathbf{\zeta})$, where  $\mathbf{\zeta} \sim \mathcal{N}(0, \sigma\mathcal{I})$ and $\sigma$  represents the noise level. 
The results in Figure \ref{fig: noise} demonstrate that PIE \textit{consistently} outperforms all the other baselines across a large range of noise levels.

\textbf{Robustness under Different Number of Input Data Points} We evaluate the robustness of our proposed method in terms of different number of input data points during evaluation. Note that the number of input data points is fixed (i.e., $N=200$) in pre-training. Results in Figure \ref{fig: noise} show that PIE is \textit{highly} robust in terms of different number of input data points.

\begin{figure}[t]
    \centering
    \subfigure[Size of dataset.]{
        \label{fig: sensitivity-pre-training-size}
    \includegraphics[width=0.23\textwidth]{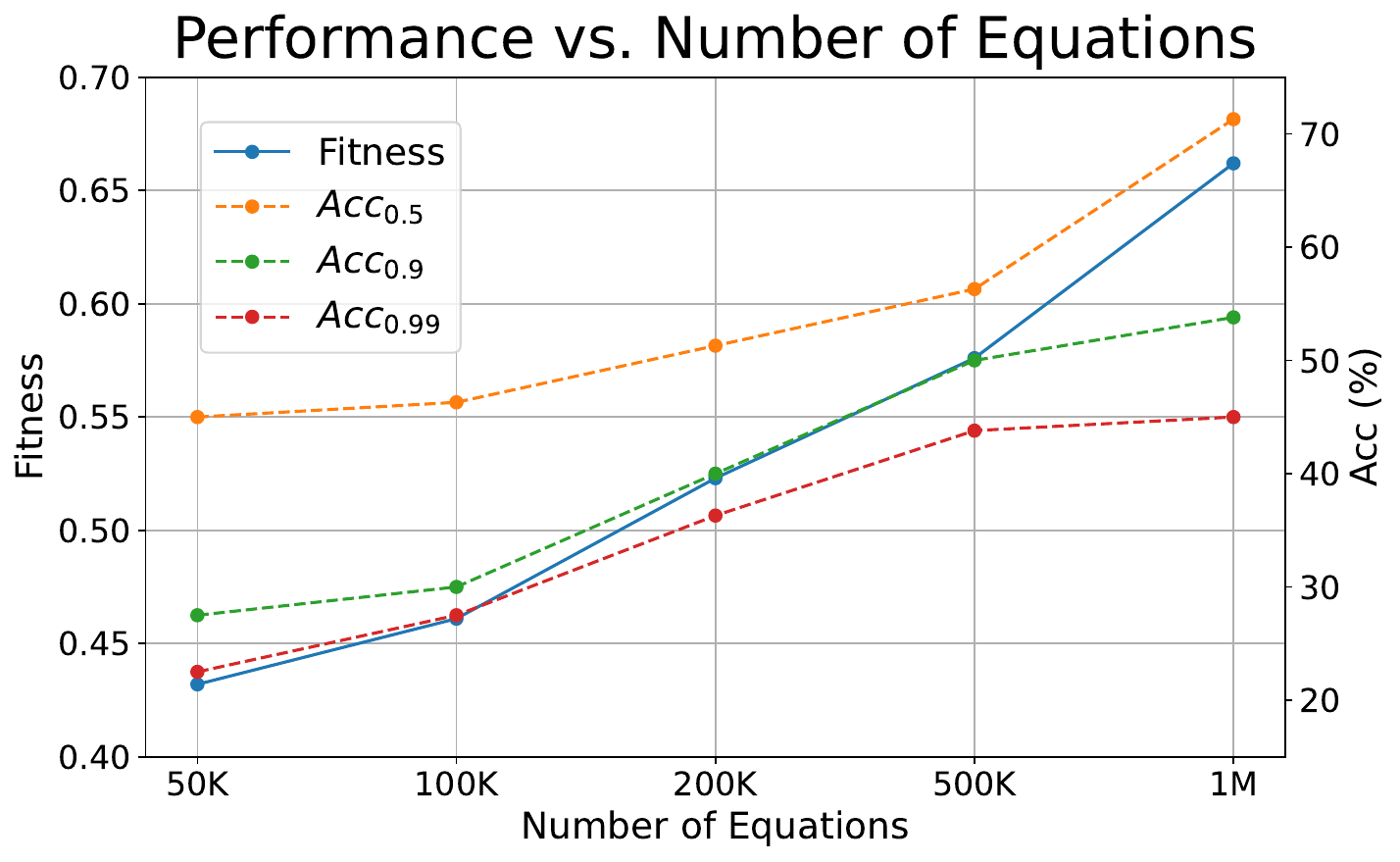}}
    \subfigure[Number of points.]{
        \label{fig: sensitivity-input-points}
    \includegraphics[width=0.23\textwidth]{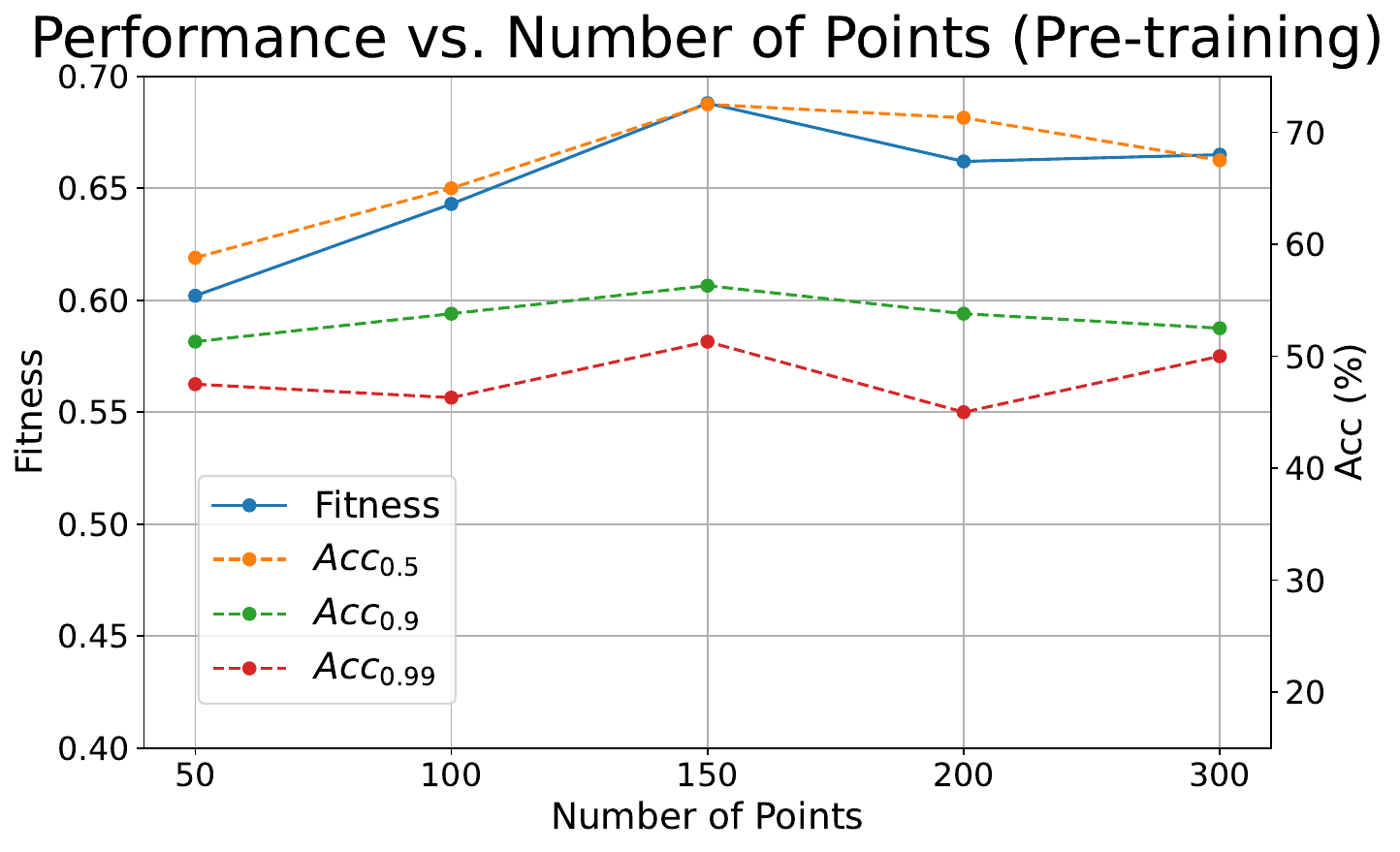}}
     \subfigure[Size of beam search.]{
        \label{fig: sensitivity-beam-search}
        \includegraphics[width=0.23\textwidth]{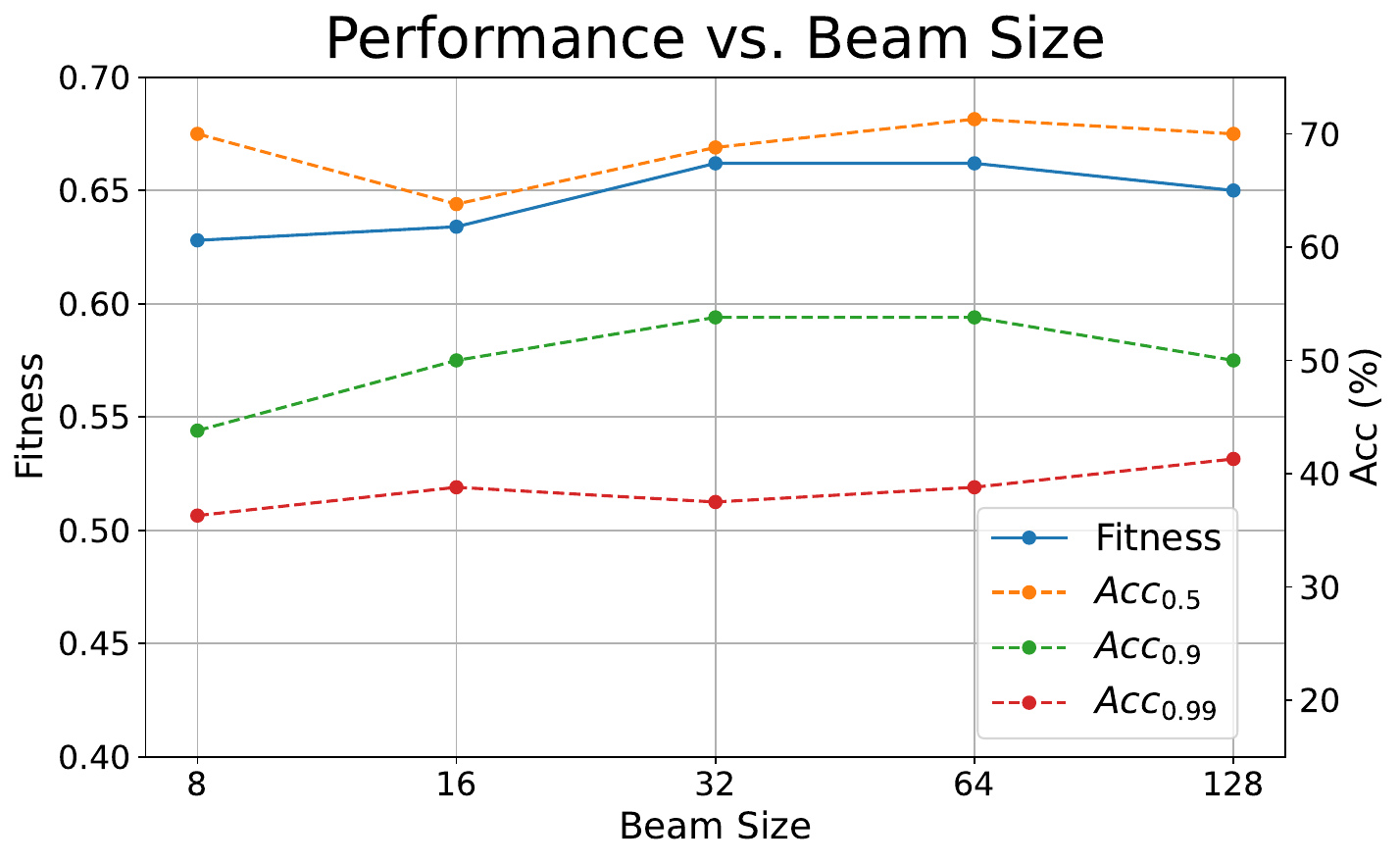}}
     \subfigure[Tolerance level.]{
        \label{fig: sensitivity-tolerance}
        \includegraphics[width=0.23\textwidth]{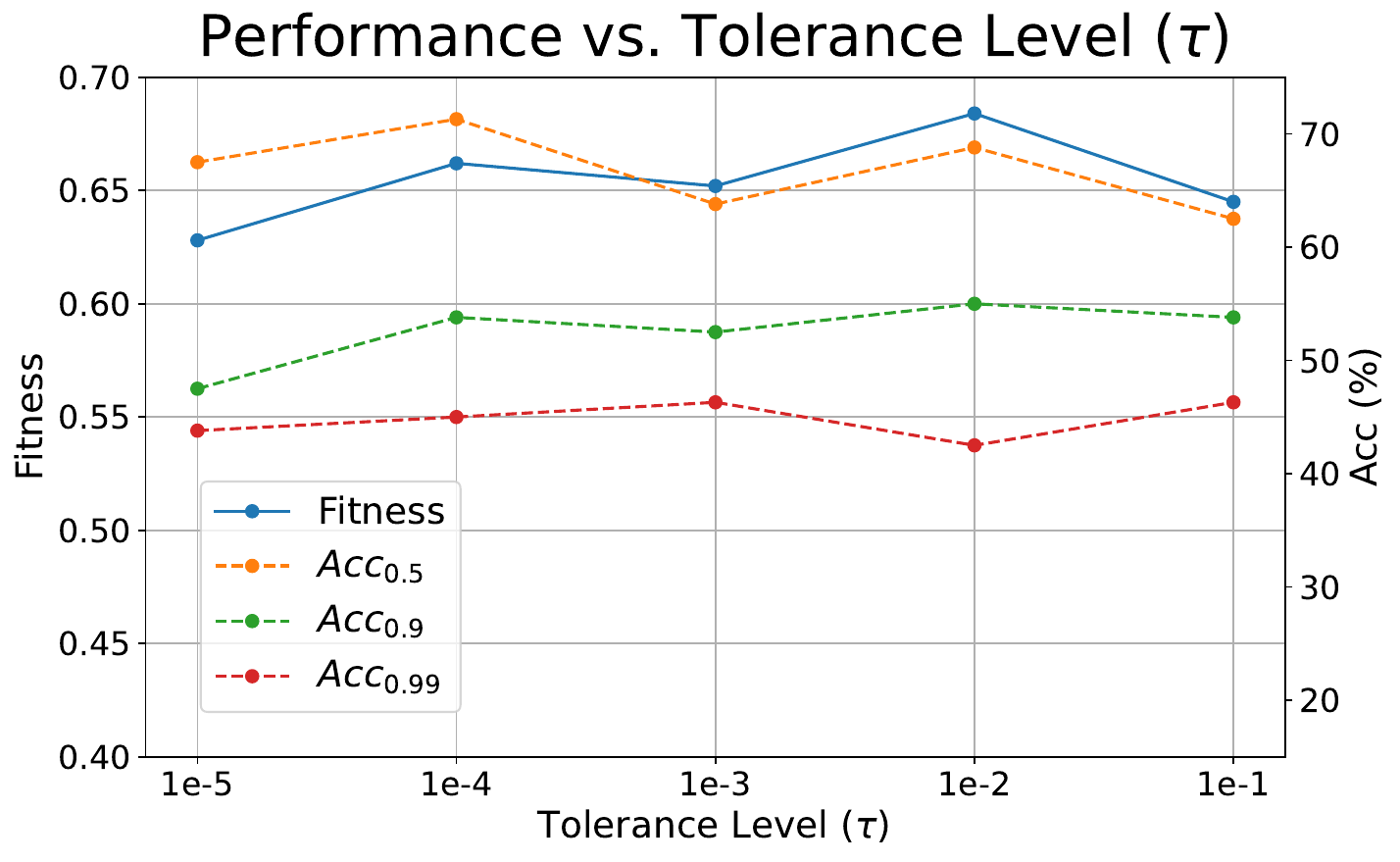}}
\vspace{-3mm}
\caption{
We evaluate different hyperparameters we used for training and inference and report their sensitivity in terms of different metrics. Results show that the performance of PIE is sensitive to the size  of the pre-training dataset, while it is relatively insensitive to the number of the input data points during training, the size of beam search during inference, and the tolerance level during inference.
}
\vspace{-5mm}
\label{fig: sensitivity}
\end{figure}

\subsection{Sensitivity Analysis}

\textbf{For training}  We evaluate PIE with different sizes $K$ of the pre-training dataset and different numbers $N$ of the input data points sampled from each implicit equation. For the latter, we evaluate PIE with fixed number of data points same to that in Section \ref{sec: comparative}. 
We report the performance with multiple metrics in Figure \ref{fig: sensitivity-pre-training-size} and Figure \ref{fig: sensitivity-input-points}. Results show that: 1) larger sizes of of the pre-training datasets lead to higher performance, as we expected; 2) either too large or too small number of the input data points lead to a degradation in performance, which might because that too large number of data points increases the learning complexity while too small makes the semantics of the input data unclear. 

\textbf{For inference} We evaluate PIE with different sizes $N_\text{BS}$ of beam search and different tolerance levels $\tau$. 
Results in Figure \ref{fig: sensitivity-beam-search} and Figure \ref{fig: sensitivity-tolerance} show that: 1) a beam search size larger than $32$ tends to achieves high performance; 2) the performance of PIE is relatively insensitive to the tolerance level $\tau$. 

\section{Conclusion} \label{sec: conclusion}
In this work, we focus on the symbolic discovery of implicit equations, which are more expressive in modeling real-world data than explicit functions. To tackle the dense distribution of degenerate equations within the search space, we propose a novel pre-training framework, PIE, to refine the search space for implicit equation discovery. Experiments show that PIE outperforms existing SR methods across in-domain accuracy, out-of-domain generalization, and robustness to noisy data.
One limitation of PIE is that, it considers a special case---i.e., the implicit equation---in unsupervised scientific discovery. 
Applying PIE to discover more general mathematical relations---e.g., data following specific distributions or parametric equations---is an exciting avenue for further research.


\begin{thebibliography}{46}
\providecommand{\natexlab}[1]{#1}
\providecommand{\url}[1]{\texttt{#1}}
\expandafter\ifx\csname urlstyle\endcsname\relax
  \providecommand{\doi}[1]{doi: #1}\else
  \providecommand{\doi}{doi: \begingroup \urlstyle{rm}\Url}\fi

\bibitem[Wang et~al.(2023)Wang, Fu, Du, Gao, Huang, Liu, Chandak, Liu, Katwyk, Deac, Anandkumar, Bergen, Gomes, Ho, Kohli, Lasenby, Leskovec, Liu, Manrai, Marks, Ramsundar, Song, Sun, Tang, Velickovic, Welling, Zhang, Coley, Bengio, and Zitnik]{ai4science}
Hanchen Wang, Tianfan Fu, Yuanqi Du, Wenhao Gao, Kexin Huang, Ziming Liu, Payal Chandak, Shengchao Liu, Peter~Van Katwyk, Andreea Deac, Anima Anandkumar, Karianne Bergen, Carla~P. Gomes, Shirley Ho, Pushmeet Kohli, Joan Lasenby, Jure Leskovec, Tie{-}Yan Liu, Arjun Manrai, Debora~S. Marks, Bharath Ramsundar, Le~Song, Jimeng Sun, Jian Tang, Petar Velickovic, Max Welling, Linfeng Zhang, Connor~W. Coley, Yoshua Bengio, and Marinka Zitnik.
\newblock Scientific discovery in the age of artificial intelligence.
\newblock \emph{Nat.}, 620\penalty0 (7972):\penalty0 47--60, 2023.
\newblock \doi{10.1038/S41586-023-06221-2}.
\newblock URL \url{https://doi.org/10.1038/s41586-023-06221-2}.

\bibitem[Schmidt and Lipson(2009{\natexlab{a}})]{db-fem}
Michael Schmidt and Hod Lipson.
\newblock Symbolic regression of implicit equations.
\newblock In \emph{Genetic programming theory and practice VII}, pages 73--85. Springer, 2009{\natexlab{a}}.

\bibitem[Chen et~al.(2018{\natexlab{a}})Chen, Zhong, and Tan]{cl-fem}
Yongliang Chen, Jinghui Zhong, and Mingkui Tan.
\newblock Comprehensive learning gene expression programming for automatic implicit equation discovery.
\newblock In \emph{Computational Science--ICCS 2018: 18th International Conference, Wuxi, China, June 11--13, 2018, Proceedings, Part I 18}, pages 114--128. Springer, 2018{\natexlab{a}}.

\bibitem[Voxman and Goetschel(2017)]{math}
William~L Voxman and Roy~H Goetschel.
\newblock \emph{Advanced Calculus: An Introduction to Modem Analysis}.
\newblock CRC Press, 2017.

\bibitem[Schmarje et~al.(2021)Schmarje, Santarossa, Schr{\"o}der, and Koch]{survey-unsupervised}
Lars Schmarje, Monty Santarossa, Simon-Martin Schr{\"o}der, and Reinhard Koch.
\newblock A survey on semi-, self-and unsupervised learning for image classification.
\newblock \emph{IEEE Access}, 9:\penalty0 82146--82168, 2021.

\bibitem[Petersen(2019)]{symb}
Brenden~K. Petersen.
\newblock Deep symbolic regression: Recovering mathematical expressions from data via policy gradients.
\newblock \emph{CoRR}, abs/1912.04871, 2019.
\newblock URL \url{http://arxiv.org/abs/1912.04871}.

\bibitem[Biggio et~al.(2021{\natexlab{a}})Biggio, Bendinelli, Neitz, Lucchi, and Parascandolo]{nesymbres}
Luca Biggio, Tommaso Bendinelli, Alexander Neitz, Aurelien Lucchi, and Giambattista Parascandolo.
\newblock Neural symbolic regression that scales.
\newblock In \emph{International Conference on Machine Learning}, pages 936--945. PMLR, 2021{\natexlab{a}}.

\bibitem[Cranmer et~al.(2020)Cranmer, Sanchez~Gonzalez, Battaglia, Xu, Cranmer, Spergel, and Ho]{symb-graph-1}
Miles Cranmer, Alvaro Sanchez~Gonzalez, Peter Battaglia, Rui Xu, Kyle Cranmer, David Spergel, and Shirley Ho.
\newblock Discovering symbolic models from deep learning with inductive biases.
\newblock \emph{Advances in Neural Information Processing Systems}, 33:\penalty0 17429--17442, 2020.

\bibitem[Shi et~al.(2022)Shi, Ding, Cao, Liu, Li, et~al.]{symb-graph-2}
Hongzhi Shi, Jingtao Ding, Yufan Cao, Li~Liu, Yong Li, et~al.
\newblock Learning symbolic models for graph-structured physical mechanism.
\newblock In \emph{The Eleventh International Conference on Learning Representations}, 2022.

\bibitem[Zhong et~al.(2022)Zhong, Yang, Chen, Liu, and Feng]{implicit-enhanced-gp}
Jinghui Zhong, Jiaquan Yang, Yongliang Chen, Wei-Li Liu, and Liang Feng.
\newblock Mining implicit equations from data using gene expression programming.
\newblock \emph{IEEE Transactions on Emerging Topics in Computing}, 10\penalty0 (2):\penalty0 1058--1074, 2022.
\newblock \doi{10.1109/TETC.2021.3068651}.

\bibitem[Udrescu and Tegmark(2020)]{ai-feynman}
Silviu-Marian Udrescu and Max Tegmark.
\newblock Ai feynman: A physics-inspired method for symbolic regression.
\newblock \emph{Science Advances}, 6\penalty0 (16):\penalty0 eaay2631, 2020.

\bibitem[Franklin and Griffiths(2014)]{franklin2014fields}
Joel Franklin and David~J Griffiths.
\newblock The fields of a charged particle in hyperbolic motion.
\newblock \emph{American Journal of Physics}, 82\penalty0 (8):\penalty0 755--763, 2014.

\bibitem[Wilson(1968)]{wilson1968kepler}
Curtis Wilson.
\newblock Kepler's derivation of the elliptical path.
\newblock \emph{Isis}, 59\penalty0 (1):\penalty0 4--25, 1968.

\bibitem[Gumerov and Duraiswami(2005)]{gumerov2005fast}
Nail~A Gumerov and Ramani Duraiswami.
\newblock \emph{Fast multipole methods for the Helmholtz equation in three dimensions}.
\newblock Elsevier, 2005.

\bibitem[Lee et~al.(2019)Lee, Lee, Kim, Kosiorek, Choi, and Teh]{set-transformer}
Juho Lee, Yoonho Lee, Jungtaek Kim, Adam Kosiorek, Seungjin Choi, and Yee~Whye Teh.
\newblock Set transformer: A framework for attention-based permutation-invariant neural networks.
\newblock In \emph{International conference on machine learning}, pages 3744--3753. PMLR, 2019.

\bibitem[Lample and Charton(2019)]{lample2019deep}
Guillaume Lample and Fran{\c{c}}ois Charton.
\newblock Deep learning for symbolic mathematics.
\newblock In \emph{International Conference on Learning Representations}, 2019.

\bibitem[Kahan(1996)]{ieee754}
William Kahan.
\newblock Ieee standard 754 for binary floating-point arithmetic.
\newblock \emph{Lecture Notes on the Status of IEEE}, 754\penalty0 (94720-1776):\penalty0 11, 1996.

\bibitem[Fletcher(2000)]{bfgs}
Roger Fletcher.
\newblock \emph{Practical methods of optimization}.
\newblock John Wiley \& Sons, 2000.

\bibitem[Koza(1994)]{koza1994genetic}
John~R Koza.
\newblock Genetic programming as a means for programming computers by natural selection.
\newblock \emph{Statistics and computing}, 4:\penalty0 87--112, 1994.

\bibitem[Brameier et~al.(2007)Brameier, Banzhaf, and Banzhaf]{brameier2007linear}
Markus Brameier, Wolfgang Banzhaf, and Wolfgang Banzhaf.
\newblock \emph{Linear genetic programming}, volume~1.
\newblock Springer, 2007.

\bibitem[Petersen et~al.(2020)Petersen, Larma, Mundhenk, Santiago, Kim, and Kim]{petersen2020deep}
Brenden~K Petersen, Mikel~Landajuela Larma, Terrell~N Mundhenk, Claudio~Prata Santiago, Soo~Kyung Kim, and Joanne~Taery Kim.
\newblock Deep symbolic regression: Recovering mathematical expressions from data via risk-seeking policy gradients.
\newblock In \emph{International Conference on Learning Representations}, 2020.

\bibitem[Kusner et~al.(2017)Kusner, Paige, and Hern{\'{a}}ndez{-}Lobato]{icml/KusnerPH17}
Matt~J. Kusner, Brooks Paige, and Jos{\'{e}}~Miguel Hern{\'{a}}ndez{-}Lobato.
\newblock Grammar variational autoencoder.
\newblock In \emph{Proceedings of the 34th International Conference on Machine Learning, {ICML} 2017}, volume~70 of \emph{Proceedings of Machine Learning Research}, pages 1945--1954. {PMLR}, 2017.

\bibitem[Popov et~al.(2023)Popov, Lazarev, Belavin, Derkach, and Ustyuzhanin]{PopovLBDU23}
Sergei Popov, Mikhail Lazarev, Vladislav Belavin, Denis Derkach, and Andrey Ustyuzhanin.
\newblock Symbolic expression generation \emph{via} variational auto-encoder.
\newblock \emph{PeerJ Comput. Sci.}, 9:\penalty0 e1241, 2023.

\bibitem[Mundhenk et~al.(2021)Mundhenk, Landajuela, Glatt, Santiago, Petersen, et~al.]{mundhenk2021symbolic}
Terrell Mundhenk, Mikel Landajuela, Ruben Glatt, Claudio~P Santiago, Brenden~K Petersen, et~al.
\newblock Symbolic regression via deep reinforcement learning enhanced genetic programming seeding.
\newblock \emph{Advances in Neural Information Processing Systems}, 34:\penalty0 24912--24923, 2021.

\bibitem[Landajuela et~al.(2022)Landajuela, Lee, Yang, Glatt, Santiago, Aravena, Mundhenk, Mulcahy, and Petersen]{landajuela2022unified}
Mikel Landajuela, Chak~Shing Lee, Jiachen Yang, Ruben Glatt, Claudio~P Santiago, Ignacio Aravena, Terrell Mundhenk, Garrett Mulcahy, and Brenden~K Petersen.
\newblock A unified framework for deep symbolic regression.
\newblock \emph{Advances in Neural Information Processing Systems}, 35:\penalty0 33985--33998, 2022.

\bibitem[Hochreiter and Schmidhuber(1997)]{hochreiter1997long}
Sepp Hochreiter and J{\"u}rgen Schmidhuber.
\newblock Long short-term memory.
\newblock \emph{Neural computation}, 9\penalty0 (8):\penalty0 1735--1780, 1997.

\bibitem[Oussidi and Elhassouny(2018)]{oussidi2018deep}
Achraf Oussidi and Azeddine Elhassouny.
\newblock Deep generative models: Survey.
\newblock In \emph{2018 International conference on intelligent systems and computer vision (ISCV)}, pages 1--8. IEEE, 2018.

\bibitem[Chowdhary and Chowdhary(2020)]{chowdhary2020natural}
KR1442 Chowdhary and KR~Chowdhary.
\newblock Natural language processing.
\newblock \emph{Fundamentals of artificial intelligence}, pages 603--649, 2020.

\bibitem[Biggio et~al.(2021{\natexlab{b}})Biggio, Bendinelli, Neitz, Lucchi, and Parascandolo]{biggio2021neural}
Luca Biggio, Tommaso Bendinelli, Alexander Neitz, Aurelien Lucchi, and Giambattista Parascandolo.
\newblock Neural symbolic regression that scales.
\newblock In \emph{International Conference on Machine Learning}, pages 936--945. Pmlr, 2021{\natexlab{b}}.

\bibitem[Valipour et~al.(2021)Valipour, You, Panju, and Ghodsi]{valipour2021symbolicgpt}
Mojtaba Valipour, Bowen You, Maysum Panju, and Ali Ghodsi.
\newblock Symbolicgpt: A generative transformer model for symbolic regression.
\newblock \emph{arXiv preprint arXiv:2106.14131}, 2021.

\bibitem[Bendinelli et~al.(2023)Bendinelli, Biggio, and Kamienny]{bendinelli2023controllable}
Tommaso Bendinelli, Luca Biggio, and Pierre-Alexandre Kamienny.
\newblock Controllable neural symbolic regression.
\newblock In \emph{International Conference on Machine Learning}, pages 2063--2077. PMLR, 2023.

\bibitem[Kamienny et~al.(2022)Kamienny, d'Ascoli, Lample, and Charton]{kamienny2022end}
Pierre-Alexandre Kamienny, St{\'e}phane d'Ascoli, Guillaume Lample, and Fran{\c{c}}ois Charton.
\newblock End-to-end symbolic regression with transformers.
\newblock \emph{Advances in Neural Information Processing Systems}, 35:\penalty0 10269--10281, 2022.

\bibitem[Li et~al.(2022)Li, Li, Sun, Wu, Yu, Liu, Li, and Tian]{li2022transformer}
Wenqiang Li, Weijun Li, Linjun Sun, Min Wu, Lina Yu, Jingyi Liu, Yanjie Li, and Songsong Tian.
\newblock Transformer-based model for symbolic regression via joint supervised learning.
\newblock In \emph{The Eleventh International Conference on Learning Representations}, 2022.

\bibitem[Holt et~al.(2023)Holt, Qian, and van~der Schaar]{holt2023deep}
Samuel Holt, Zhaozhi Qian, and Mihaela van~der Schaar.
\newblock Deep generative symbolic regression.
\newblock In \emph{The Eleventh International Conference on Learning Representations}, 2023.

\bibitem[Kamienny et~al.(2023)Kamienny, Lample, Lamprier, and Virgolin]{kamienny2023deep}
Pierre-Alexandre Kamienny, Guillaume Lample, Sylvain Lamprier, and Marco Virgolin.
\newblock Deep generative symbolic regression with monte-carlo-tree-search.
\newblock In \emph{International Conference on Machine Learning}, pages 15655--15668. PMLR, 2023.

\bibitem[Schmidt and Lipson(2009{\natexlab{b}})]{schmidt2009symbolic}
Michael Schmidt and Hod Lipson.
\newblock Symbolic regression of implicit equations.
\newblock In \emph{Genetic programming theory and practice VII}, pages 73--85. Springer, 2009{\natexlab{b}}.

\bibitem[Chen et~al.(2018{\natexlab{b}})Chen, Zhong, and Tan]{chen2018comprehensive}
Yongliang Chen, Jinghui Zhong, and Mingkui Tan.
\newblock Comprehensive learning gene expression programming for automatic implicit equation discovery.
\newblock In \emph{Computational Science--ICCS 2018: 18th International Conference, Wuxi, China, June 11--13, 2018, Proceedings, Part I 18}, pages 114--128. Springer, 2018{\natexlab{b}}.

\bibitem[Voulodimos et~al.(2018)Voulodimos, Doulamis, Doulamis, and Protopapadakis]{voulodimos2018deep}
Athanasios Voulodimos, Nikolaos Doulamis, Anastasios Doulamis, and Eftychios Protopapadakis.
\newblock Deep learning for computer vision: A brief review.
\newblock \emph{Computational intelligence and neuroscience}, 2018, 2018.

\bibitem[Jaiswal et~al.(2020)Jaiswal, Babu, Zadeh, Banerjee, and Makedon]{jaiswal2020survey}
Ashish Jaiswal, Ashwin~Ramesh Babu, Mohammad~Zaki Zadeh, Debapriya Banerjee, and Fillia Makedon.
\newblock A survey on contrastive self-supervised learning.
\newblock \emph{Technologies}, 9\penalty0 (1):\penalty0 2, 2020.

\bibitem[Goodfellow et~al.(2014)Goodfellow, Pouget-Abadie, Mirza, Xu, Warde-Farley, Ozair, Courville, and Bengio]{goodfellow2014generative}
Ian Goodfellow, Jean Pouget-Abadie, Mehdi Mirza, Bing Xu, David Warde-Farley, Sherjil Ozair, Aaron Courville, and Yoshua Bengio.
\newblock Generative adversarial nets.
\newblock \emph{Advances in neural information processing systems}, 27, 2014.

\bibitem[Kingma and Welling(2013)]{kingma2013auto}
Diederik~P Kingma and Max Welling.
\newblock Auto-encoding variational bayes.
\newblock \emph{arXiv preprint arXiv:1312.6114}, 2013.

\bibitem[Yi et~al.(2019)Yi, Walia, and Babyn]{yi2019generative}
Xin Yi, Ekta Walia, and Paul Babyn.
\newblock Generative adversarial network in medical imaging: A review.
\newblock \emph{Medical image analysis}, 58:\penalty0 101552, 2019.

\bibitem[Killoran et~al.(2017)Killoran, Lee, Delong, Duvenaud, and Frey]{killoran2017generating}
Nathan Killoran, Leo~J Lee, Andrew Delong, David Duvenaud, and Brendan~J Frey.
\newblock Generating and designing dna with deep generative models.
\newblock \emph{arXiv preprint arXiv:1712.06148}, 2017.

\bibitem[Anand and Huang(2018)]{anand2018generative}
Namrata Anand and Possu Huang.
\newblock Generative modeling for protein structures.
\newblock \emph{Advances in neural information processing systems}, 31, 2018.

\bibitem[Devlin et~al.(2018)Devlin, Chang, Lee, and Toutanova]{devlin2018bert}
Jacob Devlin, Ming-Wei Chang, Kenton Lee, and Kristina Toutanova.
\newblock Bert: Pre-training of deep bidirectional transformers for language understanding.
\newblock \emph{arXiv preprint arXiv:1810.04805}, 2018.

\bibitem[Kingma and Ba(2015)]{adam}
Diederik~P. Kingma and Jimmy Ba.
\newblock Adam: {A} method for stochastic optimization.
\newblock In Yoshua Bengio and Yann LeCun, editors, \emph{3rd International Conference on Learning Representations, {ICLR} 2015, San Diego, CA, USA, May 7-9, 2015, Conference Track Proceedings}, 2015.
\newblock URL \url{http://arxiv.org/abs/1412.6980}.

\end{thebibliography}

\clearpage
\newpage
\appendix

\section{Related Work} \label{app: related-work}

\subsection{Symbolic Regression}
\textbf{GP-based methods}\hspace{0.5em} Traditional approaches to symbolic regression are primarily based on genetic programming (GP) \cite{koza1994genetic, brameier2007linear}. GP methods try to emulate the random process of natural gene selection to search for the candidate expressions with the best fitness to the given data. Although GP-based techniques have shown potential effectiveness on function discovery, they usually learn expressions from scratch for each single task and are relatively sensitive to hyperparameters~\cite{petersen2020deep}.

\textbf{DL-based methods}\hspace{0.5em} More recently, deep learning (DL) methods for symbolic regression have been introduced to deal with high-dimensional variables and larger datasets. Some works leverage the generative model of variation autoencoder to capture the underlying distribution of data and produce the best-fitting function over the latent space~\cite{icml/KusnerPH17,PopovLBDU23}.
Some approaches explore the application of reinforcement learning (RL) on symbolic regression~\cite{petersen2020deep, mundhenk2021symbolic, landajuela2022unified}, making a significant step forward in SR methods. These methods utilize recurrent neural network (RNN) \cite{hochreiter1997long} as a policy network to model the distribution of mathematical expressions and employ a risk-seeking policy gradient to maximize the performance.
While showing promising results, these DL-based models are trained for the specific dataset and have to be retrained from scratch for each new dataset.

\textbf{Pre-training methods}\hspace{0.5em} Inspired by the success of the pre-trained large models in computer vision (CV) \cite{oussidi2018deep} and natural language processing (NLP) \cite{chowdhary2020natural}, recent studies develop pretraining-based approaches for symbolic regression to learn an end-to-end mapping from numerical dataset to the corresponding symbolic formula~\cite{biggio2021neural, valipour2021symbolicgpt, bendinelli2023controllable, kamienny2022end, li2022transformer, holt2023deep, kamienny2023deep}. These methods pre-train neural networks, such as Transformer, on a variety of datasets containing observations from different functions. After pre-training, inference is often orders of magnitude faster since no training is needed for previously unseen expressions.
However, existing SR methods are mostly focused on learning an explicit symbolic model that can describe a projection from input \(\mathbf{x}\) to the output \(y\). The use of symbolic regression remains largely unexplored for unsupervised datasets without labeling the input and output variables.

\subsection{Symbolic Discovery For Implicit Equations}
\citet{schmidt2009symbolic} pioneers the integration of implicit equations into symbolic regression research based on the GP framework. Their initial findings indicate that a simple fitness function solely minimizing $|f(x)|$ may lead to degenerate functions that are zero everywhere, i.e. both inside and outside the data distribution. To tackle this problem, they introduce a novel fitness metric, 
Derivative-Based Fitness Evaluation Mechanism (DB-FEM), which computes implicit derivatives of the given data and fits the gradients of the data to the gradients of the implicit function.
While this approach demonstrates enhanced effectiveness by incorporating gradient information into the fitness measurement, it assigns identical fitness scores to two distinct equations such as $f(\mathbf{x})$ and $f(\mathbf{x})+C$. This hampers its ability to address cases where optimization of constants in the equation is crucial. Additionally, DB-FEM suffers from high computational complexity due to the necessity of calculating partial derivatives for each variable.

\citet{chen2018comprehensive} introduces an alternative approach for discovering implicit equations through genetic programming. This work features a novel fitness metric, the Comprehensive Learning-based Fitness Evaluation Mechanism (CL-FEM), to assess the quality of implicit equations. CL-FEM detects degenerate dimensions of an implicit function by creating a synthetic dataset that evaluates whether the function's value at points outside the dataset noticeably differs from its value at points within the dataset. CL-FEM shows a significantly reduced time cost compared to DB-FEM. However, their experiments only consider simple cases that do not need to optimize constants within the equation. When constant optimization is necessary, their approach tends to yield equations that approximate zero, i.e., $f(\mathbf{x}) \approx 0$. Moreover, we observe CL-FEM still fall into approximately degenerate solutions due to the highly discrete and ill-conditioned search space (see Figure \ref{fig: examples} and Table \ref{tab: in-domain}). 
Thus, currently, few existing approaches customized for this task achieve satisfactory performance.

\subsection{Unsupervised learning for scientific discovery}
Unsupervised Learning \cite{survey-unsupervised} is a branch of machine learning that learns from unlabeled data without the necessity for explicit supervision. It has been successfully applied in many fields, such as computer vision \cite{voulodimos2018deep} and natural language processing \cite{chowdhary2020natural}. Moreover, unsupervised learning methods like generative models \cite{oussidi2018deep} and self-supervised learning \cite{jaiswal2020survey} are attracting growing interest due to their potential for scientific discovery. Generative models such as generative adversarial networks \cite{goodfellow2014generative} and variational autoencoders \cite{kingma2013auto} learn the underlying data distribution and facilitate the generation of new data from the optimized distribution. These models prove beneficial to synthesizing diverse scientific data like medical images \cite{yi2019generative}, genetic sequences \cite{killoran2017generating}, and protein functions \cite{anand2018generative}. Self-supervised learning can also be regarded as unsupervised learning as it does not rely on explicit labels. This approach enables models to learn general features through innovative strategies, such as predicting the masked part of the raw data based on the rest \cite{devlin2018bert} or employing contrastive learning \cite{jaiswal2020survey} to differentiate between similar and dissimilar data points. While most unsupervised learning methods employ deep learning models to extract meaningful representation, the complex structure of deep learning models makes them difficult to interpret and understand.

\section{Implementation Details} \label{app: implementation}

\begin{table}[t]
    \caption{Unnormalized probabilities of unary and binary mathematical operators used to the generate the pre-training dataset. }
    \centering

\begin{tabular}{ccc}
\toprule\toprule
\textbf{Operator} & \textbf{Meaning} & \textbf{Probability} \\
\midrule
add & $ + $ & 10 \\
mul & $ \times $ & 10 \\
sub & $ - $ & 5 \\
div & $ \div $ & 5 \\
sqrt & $ \sqrt{} $ & 4 \\
exp & $ \exp $ & 4 \\
ln & $ \ln $ & 4 \\
sin & $ \sin $ & 4 \\
cos & $ \cos $ & 4 \\
pow2 & $ (\cdot)^2 $ & 4 \\
pow3 & $ (\cdot)^3 $ & 2 \\
pow4 & $ (\cdot)^4 $ & 1 \\
pow5 & $ (\cdot)^5 $ & 1 \\
\bottomrule
\end{tabular}
    \label{tab: operator_probs}
\end{table}

\begin{table}[t]
\caption{The hyperparameters used in PIE. Specifically, the hyperparameters for both the sampling process and the model architecture are consistent to that in NeSymReS \citep{nesymbres}, which can be regarded as a representative of neural symbolic approaches.}
\centering
\begin{tabular}{lc}
\toprule\toprule
Parameter & Value \\   \midrule
\multicolumn{2}{c}{Implicit equation sampling:} \\
Number of non-leaf nodes & 5 \\
Probability of independent variables and constants & 0.8,0.2 \\
Size  of the pre-training dataset $K$ & 1e6 \\
Size  of the validation dataset $K$ & 1e4 \\
Size  of data points for each implicit equation $N$ & 200\\ 
Maximal dimension of input $D_{\max}$ & 3\\
\midrule
\multicolumn{2}{c}{Embedding Layer:} \\
Dimension of embedding $d_{\text{emb}}$ & 16 \\
Dimension of hidden layer $d_{\text{hid}}$ & 512 \\
\midrule
\multicolumn{2}{c}{Encoder:} \\
Dimension of hidden layer & 512 \\
Number of heads & 16 \\
Dropout rate & 0.1 \\
Number of layers & 4 \\
Number of ISABs & 5 \\
Number of PMA features & 10 \\
Number of inducing points & 50 \\
\midrule
\multicolumn{2}{c}{Decoder:} \\
Dimension of hidden layer & 512 \\
Number of heads & 16 \\
Dropout rate & 0.1 \\
Number of layers & 8 \\
Dimension of output & 60 \\
\midrule
\multicolumn{2}{c}{Training settings:} \\
Learning rate & 1e-4 \\
Optimizer & Adam \citep{adam} \\
\midrule
\multicolumn{2}{c}{Inference settings:} \\
Beam search size $N_{\text{BS}}$ & 64 \\
Tolerance level $\tau$ of CL-FEM \citep{cl-fem} & 1e-4 \\
\bottomrule
\end{tabular}
\label{tab:hyperparameters}
\end{table}
\begin{algorithm}[t]
   \caption{The Pre-training Process of PIE}
   \label{alg:pie-pre-training}
    \begin{algorithmic}[1]
        \Require {initial embedding and transformer parameters \(\theta\), batch size \(B\), training dataset \(\mathcal{D}_{\text{pre-train}}\), learning rate \(\eta\)
        }

        \While{not timeout}
            \State \(Loss(\theta)\leftarrow 0\)
            \For{\(i=1,\dots,B\)}
            \State \(\tilde{f}, \mathcal{D}\leftarrow\) sample the skeleton of an equation \(f\) and its sampled dataset form \(\mathcal{D}_{\text{pre-train}}\)
            \State \(\mathbf{z}_\text{hid}\leftarrow \texttt{EmbeddingLayer}(\mathcal{D}\mid \theta)\)
            \State \(\mathbf{z}\leftarrow \texttt{Encoder}(\mathbf{z}_\text{hid}\mid\theta)\)
            \State \(p(\tilde{f}_{l+1} \mid \tilde{f}_{1:l}, \mathbf{z})\leftarrow \texttt{Decoder}(\tilde{f}_{1:l}, \mathbf{z} \mid \theta)\left[\tilde{f}_{l+1}\right]\) for all \(l\)
            \State \(Loss(\theta)\leftarrow Loss(\theta) - \sum\limits_{l} \log p(\tilde{f}_{l+1} \mid \tilde{f}_{1:l}, \mathbf{z}) \)
            \EndFor
        \State \(\theta \leftarrow \theta - \eta \nabla Loss(\theta)\)
        \EndWhile
        \State \Return \(\theta\)
    \end{algorithmic}
\end{algorithm}

\begin{algorithm}[t]
   \caption{The Inference Process of PIE}
   \label{alg:pie-inference}
    \begin{algorithmic}[1]
        \Require {the trained embedding and transformer parameters \(\hat{\theta}\), sampled dataset \(\mathcal{D}\) from an unknown implicit function, beam search size \(N_\text{BS}\)
        }
        
        \State \(\mathbf{z}_\text{hid}\leftarrow \texttt{EmbeddingLayer}(\mathcal{D}\mid \hat{\theta})\)
        \State \(\mathbf{z}\leftarrow \texttt{Encoder}(\mathbf{z}_\text{hid}\mid\hat{\theta})\)
        \State sequentially generate \(N_\text{BS}\) skeleton candidates \(\bar{f}_k\) using beam search, scored by \(\texttt{Decoder}(\cdot,\mathbf{z} \mid \hat{\theta})\) 
        \State optimize the fitness function~\eqref{eq: cl-fem} using BFGS to recover the constants of each skeleton \(\bar{f}_k\)
        \State select the expression \(f\) with the highest fitness score from the \(N_\text{BS}\) candidates
        \State \Return \(f\)
    \end{algorithmic}
\end{algorithm}

\vspace{-0.25cm}
 
\textbf{Training and Inference} We use the Adam \cite{adam} optimizer for model optimization. 
Each mini-batch includes $64$ training samples. The learning rate is fixed at $10^{-4}$ and no learning rate scheduler is applied during training. To evaluate the model performance during training, we use a validation set consisting of $10,000$ samples. If the model's performance does not improve on the validation set for one epoch, we early stop the training procedure to prevent overfitting. During inference, we utilize a beam search strategy with $N_\text{BS}=64$ to identify the ground truth implicit equation. When optimizing the constants in the implicit equation, we employ the BFGS algorithm without restarting and set the  tolerance level of CL-FEM as $\tau=10^{-4}$. We conduct our experiments on a machine with 2 Intel(R) Xeon(R) CPU E5-2667 v4 @ 3.20GHz and 8 Nvidia RTX 2080 GPUs, each with 12GB of memory. We spend approximately 20 hours generating the dataset used for pre-training and about 7 hours pre-training our model.
We report all the hyperparameters we used in Table \ref{tab:hyperparameters}

\textbf{Metric} We provide a detailed explanation of the metric implementation, specifically illustrating how to calculate the $MSE$ and the $NMSE$ as defined by Equations \ref{mse} and \ref{nmse}, respectively. 

To calculate the $MSE$ from Equation \ref{mse}, we proceed as follows:
1) randomly sample $M=200$ points $\{\mathbf{\hat{x}}_i\}_{i=1}^{200}$ from standard normal distribution $\mathcal{N}(\mathbf{0}, \mathcal{I})$ as initial solutions, 2) generate $M$ solutions $\{\mathbf{x}_i\}_{i=1}^{200}$ by solving the equation $\hat{f}(\mathbf{x})=0$ using these initial solutions, 3) compute the $MSE$ as the mean square error $\frac{1}{M}\sum_{i=1}^{200}|f(\mathbf{x}_i)|^2$. 

For the $NMSE$ from Equation \ref{nmse}, the steps are: 1) randomly sample $D$ points $\{\mathbf{\tilde{x}}_j\}_{j=1}^D$ from standard normal distribution $\mathcal{N}(\mathbf{0}, \mathcal{I})$. 2) calculate the normalized term $\frac{1}{D}\sum_{j=1}^D|f(\mathbf{\tilde{x}}_j)|^2$. 3) Divide the $MSE$ by the normalization term to obtain the $NMSE$. During our experiments, we varied $D$ from $10$ to $1000$ and found no significant influence on the normalization term. Consequently, we select $D=10$ as our choice for fast sampling.

\section{Baselines}\label{app: baseline}

\begin{table}[t]
\centering
\caption{Hyperparameters for GP.}
\begin{tabular}{lc}
\toprule\toprule
Parameter name        & Value \\ \midrule
population size       & 2000  \\ 
generations           & 20    \\ 
tournament size       & 20    \\ 
crossover probability           & 0.9 \\
subtree mutation probability    & 0.01\\
\bottomrule
\end{tabular}
\label{tab: gp}
\end{table}

\begin{table}[t]
\centering
\label{tab: nggp}
\caption{Hyperparameters for NGGP.}
\begin{tabular}{lc}
\toprule\toprule
Parameter name        & Value \\ \midrule
Entropy coefficient \(\lambda_H\) & 0.03  \\ 
Risk factor \(\epsilon\)          & 0.05 \\ 
Generations           & 20    \\ 
Crossover probability & 0.5   \\ 
Mutation probability  & 0.5   \\ 
number of samples   & 200000    \\
priority queue training & True  \\
\bottomrule
\end{tabular}
\end{table}

\textbf{Genetic Programming (GP)}.\hspace{0.5em} For Genetic Programming
, we utilize the open-source Python library, PySR\footnote{\url{https://gplearn.readthedocs.io/en/stable/}}, for Genetic Programming tasks. The hyper-parameters selected for our experiments are detailed in Table \ref{tab: gp}. The default set of symbols that the algorithm can use to construct mathematical expressions is given by: \(\mathcal{L} = \{+, -, \times, \div, \text{Pow}, \sqrt, \exp, \log, \text{neg}, \sin, \cos\}\)

\textbf{Neural Guided Genetic Programming (NGGP)}.\hspace{0.5em} For NGGP, we adopt the standard parameter settings provided by the open-source implementation\footnote{\url{https://github.com/dso-org/deep-symbolic-optimization}}. There are two critical hyper-parameters in NGGP:  the entropy coefficient \(\lambda_H\) and the risk factor \(\epsilon\),NGGP depends on two main hyper-parameters namely the entropy coefficient \(\lambda_H\) and the risk factor \(\epsilon\), in addition to parameters related to hybrid methods of genetic programming. The entropy coefficient \(\lambda_H\) assigns a bonus, proportional to the entropy of the sampled expression, to the primary objective. The adjustment in the formulation of the final objective is influenced by the \( (1 - \epsilon) \) quantile of the reward distribution under the current policy. The hyperparameters are listed in Table \ref{tab: nggp}.

\section{Benchmarks}\label{app: benchmark}

\begin{table}[t]
    \centering
    \caption{AI-Feynman equation with 2 to 3 independent variables.}

    \begin{tabular}{ll}
    \toprule\toprule
    Expression & Expression \\
    \midrule
    $x_1 x_2 - 0.564$ & $x_1 x_2 x_3 - 0.824$ \\
    $\frac{x_1 x_2^2}{2} - 0.512$ & $0.0796 \frac{x_1}{x_2 x_3^2} - 0.817$ \\

    $-0.238 + \frac{x_2 + x_3}{1 + \frac{x_2 x_3}{x_1^2}}$ & $\frac{x_1}{\sqrt{1 - \frac{x_2^2}{x_3^2}}} - 0.316$ \\

    $\frac{x_1}{x_2} - 1.909$ & $\frac{x_1 x_2}{\sqrt{1 - \frac{x_2^2}{x_3^2}}} - 1.1$ \\

    $\frac{x_1 x_2}{\sqrt{1 - \frac{x_2^2}{x_3^2}}} - 1.1$ & $x_1 x_2 \sin(x_3) - 0.075$ \\

    $-0.192 + \frac{0.399 \exp\left(-\frac{(x_2 - x_3)^2}{2 x_1^2}\right)}{x_1}$ & $-1.98 + \frac{0.399 \exp\left(-\frac{x_2^2}{2 x_1^2}\right)}{x_1}$ \\

    $\frac{x_1 \sin^2\left(\frac{x_2 x_3}{2}\right)}{\sin^2\left(\frac{x_2}{2}\right)} - 0.207$ & $0.159 x_1 x_2 - 0.337$ \\

    $\frac{3 x_1 x_2}{2} - 0.582$ & $\frac{x_3}{1 - \frac{x_2}{x_1}} - 0.772$ \\

    $\sqrt{\frac{x_1 x_2}{x_3}} - 0.994$ & $x_3 \left(1 + \frac{x_2}{x_1}\right) \sqrt{\frac{1}{1 - \frac{x_2^2}{x_1^2}}} - 0.266$ \\

    $0.0796 \frac{x_1}{x_2^2} - 0.005$ & $x_1 x_2 x_3 - 1.467$ \\

    $0.0796 \frac{x_1}{x_2 x_3} - 1.143$ & $\frac{x_1 x_3^2}{\sqrt{1 - \frac{x_2^2}{x_3^2}}} - 1.95$ \\

    $\frac{x_2 x_3}{x_1 - 1} - 0.824$ & $x_1 x_2^2 - 1.38$ \\

    $\frac{x_1 x_2^2}{2} - 1.751$ & $\frac{x_1}{x_2 (x_3 + 1)} - 0.507$ \\

    $x_1 x_2 x_3^2 - 1.454$ & $\frac{x_1 x_2}{-\frac{x_1 x_2}{3} + 1} - 0.762$ \\

    $\frac{x_1 x_2 x_3}{2} - 1.84$ & $\frac{0.159 x_1 x_2}{x_3} - 0.948$ \\

    $0.0477 \frac{x_1^2}{x_2 x_3} - 0.819$ & $\frac{12.568 x_1 x_2}{x_3} - 0.726$ \\

    $0.159 x_1 x_2 - 1.703$ & $0.0796 \frac{x_1 x_2}{x_3} - 0.279$ \\

    $\frac{x_1}{2 x_2 + 2} - 0.389$ & $2 x_1 (1 - \cos(x_2 x_3)) - 1.361$ \\

    $6.284 \frac{x_1}{x_2 x_3} - 1.753$ &   $x_1 (x_2 \cos(x_3) + 1) - 0.324$ \\

    $0.0127 \frac{x_1^2}{x_2 x_3^2} - 1.343$    &   $\sqrt{\frac{x_1 x_2}{x_3}} - 1.643$ \\

    \bottomrule
    \end{tabular}
    \label{tab: aif}
\end{table}

\begin{table}[t]
    \centering
    \caption{$50$ equations randomly extracted from the Synthetic dataset.}
    \begin{tabular}{ll}
    \toprule\toprule
    Expression & Expression \\
    \midrule
    $x_1 + x_2(x_2 + x_3) + 0.612$ & $x_2\cos(x_1^2) - 1.127$ \\
    $0.697x_1 - 1.127 + \frac{(0.612 - x_2)}{x_3}$ & $x_1 - \frac{x_1}{x_2^2} + 0.342$ \\
    $x_2\sin(x_1) + 0.612$ & $2x_1 + 2x_2 - 0.259$ \\
    $\sin(x_1) - \cos(x_2) + 0.612$ & $\log(-1.913\sin(\frac{x_1}{x_2})) + 0.608$ \\
    $0.542*x_1*\exp(\exp(x_2)) - 0.987$ & $\sin(0.779x_1\frac{(x_3 - 2.669)}{x_2}) + 0.185$ \\
    $2x_2 + \log(x_1) + 0.612$ & $x_1^{0.522} + x_1x_2 + 0.268$ \\
    $x_1 + (x_2 - 0.554)\exp(x_3) + 0.768$ & $-0.010x_1x_2^{-0.090} + x_1 + 0.170$ \\
    $\frac{x_1}{sin(x_2 - 0.534)} - x_3 - 2.766$ & $14.590(x_1 + 0.076x_2)^{3.298} - 1.327$ \\
    $-0.417 - 0.992\frac{(x_1 + 2x_2)}{x_3}$ & $-5.592x_1\cos(x_2) + 0.015x_1 + 0.883$ \\
    $2x_2 + 0.987\exp(x_1) + 0.534$ & $x_1 + x_3 + \exp(x_2) - 2.622$ \\
    $x_2\sin(x_1) + 0.308$ & $x_1 + x_2 - 1.309$ \\
    $x_1 + x_2\exp(x_3) - 1.595$ & $x_1 + x_2 + \exp(x_1) - 0.263$ \\
    $-0.534x_1 + 0.707\sqrt{-x_1*x_2} - 3.793$ & $x_1 + x_2 + 0.484$ \\
    $-0.174 + 0.917\frac{(x_1 + 0.941x_2)}{x_3}$ & $-0.148 + x_3\frac{(x_1 - 0.107x_2)}{x_2}$ \\
    $\frac{x_1^{1.974}}{x_2} + 0.612$ & $x_1^3x_2 - 2.566$ \\
    $x_2^2(x_1 - 0.675) + x_2 + 0.773$ & $0.342 + x_3\frac{(x_1 - x_2)}{x_2}$ \\
    $-x_3 + \frac{(x_1 + 1.649)}{\sin(x_2 + 0.182)} - 0.171$ & $\frac{x_1}{(x_2 + 0.497)} - 0.665$ \\
    $0.480(x_1 - 1.855)(x_1 + x_2^2 + 1.146) + 1.249$ & $x_1^2 + x_3(x_2 + 0.292) - 2.533$ \\
    $\sqrt{x_1 - 0.612}\exp(-x_2/2) - 0.987$ & $\sin(x_2\exp(x_1)) - 0.531$ \\
    $0.500x_2 + \exp(x_1) + 2.540$ & $x_2 + \exp(x_1/2) - 0.148$ \\
    $2x_1 + 25.333x_2 - 0.174$ & $1.285x_1^3x_2 - 0.532$ \\
    $x_2\cos(x_1^2) + 1.145$ & $(x_1 - 1.262)(2x_1 - x_2) + 0.500$ \\
    $\sin(x_1\frac{(x_3 + 0.410)}{(x_2 - 2.678)}) + 0.051$ & $0.808x_2 + \exp(0.297x_1 + 0.266) + 1.138$ \\
    $x_1x_2^2 + x_2 - 1.327$ & $(x_1 - 0.791)^2(-x_1 + x_2 - 1.769) - 0.941$ \\
    $x_1(-0.137x_1 + x_2^2) + 0.757$ & $x_1 - 0.755x_2 - 1.228$ \\
    \bottomrule
    \end{tabular}
    \label{tab: synthetic}
\end{table}

This section describes the exact implicit equations utilized for benchmarking our method against the state-of-the-art baselines. The equations in the AIF dataset are detailed in Table \ref{tab: aif}. Similarly, a set of 50 equations, randomly extracted from the Synthetic dataset, is presented in Table \ref{tab: synthetic}.

\clearpage
\newpage

\end{document}